\documentclass[fleqn,usenatbib]{mnras}

\usepackage{newtxtext,newtxmath}

\usepackage[T1]{fontenc}

\DeclareRobustCommand{\VAN}[3]{#2}
\let\VANthebibliography\thebibliography
\def\thebibliography{\DeclareRobustCommand{\VAN}[3]{##3}\VANthebibliography}

\usepackage{graphicx}	
\usepackage{amsmath}	
\usepackage{subcaption}
\usepackage{cleveref}
\usepackage{tabularx}

\graphicspath{{./}{figures/}}

\newcommand{\cha}{{\it Chandra}}
\newcommand{\Msun}{${\rm M}_{\odot}$}
\newcommand{\src}{V4641\,Sgr}
\newcommand{\sw}{{\it Swift}}
\newcommand{\nic}{{\it NICER}}
\newcommand{\nus}{{\it NuSTAR}}
\newcommand{\lx}{$L_{X}$}
\newcommand{\ledd}{$L_{\rm Edd}$}

\title[High resolution X-ray spectroscopy of V4641 Sgr]{High resolution X-ray spectroscopy of V4641 Sgr during its 2020 outburst}
\author[A. W. Shaw et al.]{A. W. Shaw,$^{1}$\thanks{E-mail: aarrans@unr.edu}
J. M. Miller,$^{2}$
V. Grinberg,$^{3}$
D. J. K. Buisson,$^{4}$
C. O. Heinke,$^{5}$
R. M. Plotkin,$^{1}$
\newauthor J. A. Tomsick,$^{6}$
A. Bahramian,$^{7}$
P. Gandhi,$^{4}$
and G. R. Sivakoff$^{5}$
\\
$^{1}$Department of Physics, University of Nevada, Reno, NV 89557, USA\\
$^{2}$Department of Astronomy, University of Michigan, 1085 South University Avenue, Ann Arbor, MI 48109-1107, USA\\
$^{3}$European Space Agency (ESA), European Space Research and Technology Centre (ESTEC), Keplerlaan 1, 2201 AZ Noordwijk, The Netherlands\\
$^{4}$School of Physics and Astronomy, University of Southampton, Highfield, Southampton SO17 1BJ, UK\\
$^{5}$Department of Physics, University of Alberta, CCIS 4-181, Edmonton, AB T6G 2E1, Canada\\
$^{6}$Space Sciences Laboratory, 7 Gauss Way, University of California, Berkeley, CA 94720-7450, USA\\
$^{7}$International Centre for Radio Astronomy Research -- Curtin University, GPO Box U1987, Perth, WA 6845, Australia\\
}

\date{Accepted 2022 August 2. Received 2022 July 22; in original form 2022 June 6}

\pubyear{2022}

\begin{document}
\label{firstpage}
\pagerange{\pageref{firstpage}--\pageref{lastpage}}
\maketitle

\begin{abstract}
We observed the Galactic black hole X-ray binary \src\ with the high resolution transmission gratings on \cha\ during the source's 2020 outburst. Over two epochs of \cha\ gratings observations, we see numerous highly ionized metal lines, superimposed on a hot, disc-dominated X-ray continuum. The measured inner disc temperatures and luminosities imply an unfeasibly small inner disc radius, such that we suggest that the central engine of \src\ is obscured, and we are viewing scattered X-rays. We find that the emission lines in the \cha\ spectra cannot be constrained by a single photoionized model, instead finding that two separate photoionized model components are required, one to reproduce the iron lines and a second for the other metals. We compare the observed X-ray spectra of \src\ to optical studies during previous outbursts of the source, suggesting that the lines originate in an accretion disc wind, potentially with a spherical geometry.
\end{abstract}

\begin{keywords}
accretion, accretion discs -- X-rays: binaries -- stars: black holes -- X-rays: individual: V4641~Sgr -- line: identification -- plasmas
\end{keywords}



\section{Introduction} 
\label{sec:intro}

Galactic black hole transients are typically low-mass X-ray binaries (LMXBs) in which a black hole (BH) accretes material from a donor star via an accretion disc. They spend long periods of time in a quiescent state, characterized by low X-ray luminosities \lx$\lesssim10^{-5}L_{\rm Edd}$ (where \ledd\ is the Eddington luminosity) and an X-ray spectrum consistent with a power law with photon index $\Gamma\sim2$ \citep[e.g.][]{Reynolds-2013,Plotkin-2013}. Quiescence is followed by bright outbursts during which the source luminosity increases by several orders of magnitude across most of the electromagnetic spectrum \citep[see e.g.][]{Remillard-2006,TetarenkoB-2016}. During an outburst a BH-LMXB will transition through several `accretion states,' (typically) identified by, and named after, the properties of the X-ray spectrum \citep[see e.g.][for reviews]{Done-2007,Motta-2021}. Due to correlated timing properties, variability can also be used to identify the accretion states. Typical X-ray binary outbursts evolve from the quiescent state to a hard accretion state (and vice versa), with a large fraction of outbursts showing (potentially multiple)  transitions between the hard and soft accretion state. The soft state is characterized by a multi-temperature disc-dominated X-ray spectrum and little to no short-term variability. Conversely, the X-ray spectrum in the hard state is usually well described by a power law with $\Gamma\sim1.7$ and the transition from (to) the hard state is often marked with the disappearance (emergence) of a compact radio jet and the appearance (disappearance) of strong, fast variability in the X-ray and optical light curves \citep[e.g.][]{Belloni-2005,Remillard-2006,Gandhi-2016}.

The soft state is often strongly associated with the emergence of strong outflows in the form of accretion disc winds,  seen at X-ray -- near-infrared wavelengths. So far,  X-ray winds have only been detected in high inclination systems, which suggests they have an equatorial geometry \citep{Diaz-Trigo-2006,Miller-2006a,Ponti-2012}. In addition, until recently, it was assumed that the presence of an X-ray wind was linked to the spectral state, appearing only when the jet was quenched in the soft state \citep{Miller-2006a,Neilsen-2009,Ponti-2012,Ponti-2016}. However, the hard-state outburst of V404 Cyg in 2015 showed that winds can, in fact, be seen simultaneously with the jet \citep{Munoz-Darias-2016,King-2015}, further complicating already complex theories of accretion-outflow coupling. \citet{Homan-2016} showed that hard state winds can also be seen in neutron star LMXBs \citep[see also][]{CastroSegura-2022}.

Accretion disc winds typically manifest in the X-ray spectrum as blue-shifted H- and He-like absorption lines  \citep[e.g.][]{Lee-2002,Ueda-2004,Miller-2006a,Miller-2008b}. However, some sources exhibit strong emission lines with associated \mbox{P-Cygni} profiles \citep{Brandt-2000,King-2015}, which are also seen in optical spectra \citep{Munoz-Darias-2016,Munoz-Darias-2018} and are a sign of photons being scattered out of our line of sight, potentially even hinting at a spherical geometry. Studies of accretion disc winds through optical and (high-resolution) X-ray spectroscopy allow us to investigate the potential launching mechanisms \citep[see][and references therein]{Miller-2008b}.


\section{V4641 Sgr}
\label{sec:source}

\src\ (SAX\,J1819.3$-$2525) is an X-ray binary (often classified as an {\em Intermediate}-mass X-ray binary) containing a $6.4\pm0.6$ \Msun\ BH accreting matter from a $2.9\pm0.4$ \Msun\ B9III companion \citep{Macdonald-2014} with an orbital period $P_{\rm orb}=2.82$ days \citep{Orosz-2001}. \citet{Macdonald-2014} determined the distance to \src\ to be $d=6.2\pm0.7$ kpc, consistent with the distance distribution calculated using the {\em Gaia} data release 2 \citep[DR2;][]{Gaia-2018a} parallax assuming a volume density prior for LMXBs in the Milky Way \citep{Gandhi-2019,Atri-2019}. 

The outburst characteristics of \src\ are somewhat unusual compared to other LMXBs. 
It exhibited a 12.2 Crab outburst in September 1999 that lasted only $<2$ hours before fading to 0.1 Crab \citep{Smith-1999,Hjellming-2000}; this extremely bright outburst was also much more rapid than those of typical LMXBs.
\src\ has since shown regular outbursts with a median recurrence time of $\sim220$ days \citep{TetarenkoB-2016}, but none have reached the extreme luminosity of the 1999 event. Another unusual aspect of the system is that it has been seen to remain in a soft, disc-dominated state at very low Eddington fractions, exhibiting soft X-ray spectra at an X-ray luminosity $L_{X}\sim0.01$ \ledd\ during the 2014 outburst \citep{Pahari-2015} and as low as $L_{X}\sim6\times10^{-4}$ \ledd\ during the 2015 outburst \citep{Bahramian-2015a}. \src\ finds itself in a growing population of BH-LMXBs \citep[see also XTE\,J1720-318  4U\,1630-47 and Swift J1753.5-0127;][]{Kalemci-2013,Tomsick-2014,Shaw-2016b} that appear to remain in soft states at lower Eddington fractions than the average soft-to-hard transition luminosity of $L_{X}\sim0.03$ \ledd\ \citep[][]{Dunn-2010}.

In this work we present the first ever high-resolution X-ray spectroscopic study of \src\, obtained with the \cha\ High Energy Transmission Grating Spectrometer at two epochs during its 2020 outburst. In Section \ref{sec:obs} we present our observations and the data reduction procedures. We then present our spectral analysis in Section \ref{sec:results}, including a detailed look at the lines found in the \cha\ data. We discuss our results in \ref{sec:discussion}, performing plasma diagnostics with the observed emission lines and fitting photoionization models to the X-ray spectra. Finally, we look at the future of high-resolution X-ray spectroscopy for studies like this one in Section \ref{sec:future} and present our conclusions in Section \ref{sec:conclusions}.

\section{Observations and Data Reduction}
\label{sec:obs}

Increased X-ray activity from \src\ was first detected by the Gas Slit Camera instrument on the Monitor for All-Sky X-ray Image Gas Slit Camera \citep[{\em MAXI}/GSC;][]{Matsuoka-2009} on 2020 Jan 7, when the source emerged from the Sun constraint \citep{Shaw-2020a}. It is likely that the outburst commenced prior to the January detection but the source was obscured by the Sun. Follow-up observations were initially limited due to the Sun constraint. However, as the outburst progressed, \src\ became accessible to more observatories. Here we detail the X-ray observations and data reduction performed.

\subsection{X-ray monitoring}
\label{sec:monitoring}

In addition to the {\em MAXI}/GSC monitoring, regular pointed X-ray observations of \src\ were performed with the X-ray Telescope on-board the Neil Gehrels \sw\ Observatory \citep[\sw/XRT;][]{Burrows-2005} and the X-ray Timing Instrument on the Neutron star Interior Composition Explorer \citep[\nic/XTI;][]{Gendreau-2016}.

\sw/XRT monitored \src\ from 2020 Feb 11 -- 2020 Apr 19 (MJD 58890 -- 58958; ObsID 00013205001 -- 00013205016). Most observations were performed in windowed timing (WT) mode. However, the observation on 2020 Apr 7 (MJD 58946) was performed in photon counting (PC) mode owing to the low count rate (0.18 count s$^{-1}$). Two more observations on 2020 Apr 18 -- 19 (MJD 58957 -- 58958) were performed in both PC and WT mode, but the PC mode observations were heavily piled up, so we only utilized the WT mode observations on those dates. Data were reprocessed using the {\tt xrtpipeline} tool, part of {\sc HEAsoft} v6.28 software suite of analysis tools for high energy astronomical data\footnote{\href{https://heasarc.gsfc.nasa.gov/docs/software/heasoft/}{https://heasarc.gsfc.nasa.gov/docs/software/heasoft/}}. Spectra and associated response files were generated using the {\tt xrtproducts} tool, with source photons extracted from a circular region with a 20 pixel ($\sim47$\arcsec) radius. Background counts were extracted from an annulus centered on the source, with inner and outer radii 80 and 120 pixels, respectively. Response matrices were generated using version 20200724 of the calibration database ({\sc caldb}). Spectra were grouped such that each bin contained a minimum of 15 counts.

\nic/XTI monitored \src\ regularly throughout the first half of 2020. We choose observations that cover a similar time frame to the \sw\ coverage, in the range 2020 Jan 31 -- 2020 May 4 (MJD 58879 -- 58973; ObsIDs 3200300101 -- 3200300129). Data were processed using {\tt nicerl2} and spectra extracted using {\sc xselect}. Canned response matrices were obtained from the {\sc caldb} and background spectra were created using the {\tt nibkgestimator} tool. Similar to \sw\, \nic\ spectra were grouped such that each spectral bin contained a minimum of 15 counts

\subsection{\nus}
\label{sec:nus}

At the beginning of the outburst, \src\ was Sun-constrained to most imaging X-ray telescopes. However, the Nuclear Spectroscopic Telescope Array \citep[\nus\;][]{Harrison-2013} can point close to, and even at, the Sun at the expense of accurate pointing reconstruction. Therefore, we obtained Director's Discretionary Time (DDT) observations of \src\ with \nus\ on 2020 Jan 22 for a total exposure time of 27.8 ks. Data were reduced using the \nus\ data analysis software (NuSTARDAS) v2.0.0 originally packaged with {\sc HEAsoft} v6.28.  The {\tt nupipeline} tool was used to perform standard reduction tasks, including filtering for high levels of background during the telescope's passage through the South Atlantic Anomaly.

Due to the near-solar pointing of \nus\ during the observation, the majority of the observation occurred in observation mode 06. Mode 06 is utilized when an aspect solution is not available from the on-board star tracker located on the X-ray optics bench (Camera Head Unit \#4; CHU4). In such a scenario, the aspect reconstruction is derived using CHUs 1, 2 and 3 with degraded positional accuracy. This mode leads to images with multiple centroids from the same target, often manifesting as a seemingly elongated source. To allow accurate scientific analysis we used the {\tt nusplitsc} tool to decompose the mode 06 image into four images for each of Focal Plane Modules A and B (FPMA and FPMB), representative of the four CHU combinations \nus\ used.

Approximately 3.4 ks of the observation was performed in SCIENCE mode, also known as `mode 01,' the standard operating mode of \nus. We extracted source and background spectra and requisite response files using {\tt nuproducts}. Source spectra were extracted using a circular source extraction region with radius 70\arcsec\ and background spectra were extracted from a 90\arcsec\ radius circular region centered on a source-free region of the same chip as the target.

The remaining exposure was performed in mode 06 and, though the method of spectral extraction was the same, 
we only extracted data from mode 06 images in which the source exhibited a circular PSF, namely CHU2\footnote{For CHU2, we had to invoke the splitmode=STRICT and timecut=yes flags in {\tt nusplitsc} to produce an image with a circular PSF} and CHU3. Source spectra were extracted from a circular region with radius 60\arcsec\ for all images, while background spectra were 
extracted from a 70\arcsec\ radius circular region from the CHU2 images and a 90\arcsec\ radius circular region for CHU3.\footnote{ The reason for the discrepancy between background region sizes is due to the source in the CHU2 image having a larger PSF than in the CHU3 image. We therefore reduced the size of the background region in CHU2 to avoid source photons from the wings of the CHU2 PSF.} The total useful exposure time for each focal plane module, using mode 01 and 06 data, was 7.5\,ks. For each focal plane module we had three spectra (mode 01, CHU2 and CHU3), which were grouped using the {\tt ftgrouppha} tool, such that each spectral bin had a minimum S/N=5.

\subsection{\cha}
\label{sec:HETG}

\cha\ observed \src\ on 2020 Feb 14 and 15 (MJD 58893 and 58894) for 44.0 and 29.4 ks, respectively (ObsIDs 22389 and 23158; PI: Shaw). We used the High Energy Transmission Grating Spectrometer \citep[HETGS;][]{Canizares-2005}, which uses two gratings, the High Energy Grating (HEG) and Medium Energy Grating (MEG), to disperse photons across the Advanced CCD Imaging Spectrometer \citep[ACIS;][]{Garmire-2003} S chips.

Data were reduced and spectra were extracted using the \cha\ Interactive Analysis of Observations software \citep[CIAO;][]{Fruscione-2006} v4.12. Data were mostly reduced following standard procedures, using the {\tt chandra\_repro} tool, following CIAO threads and using CIAO CALDB v4.9.1. However, we used slightly narrower spectral extraction regions than the default to reduce overlap between the two gratings and to improve spectral S/N at shorter wavelengths (higher energies). The average \cha\ count rate in Epoch 1 was measured to be 3.5 count s$^{-1}$ and 6.0 count s$^{-1}$ in the HEG and MEG, respectively, while in Epoch 2 we measured 1.9 count s$^{-1}$ and 3.1 count s$^{-1}$ in HEG and MEG, respectively. The \cha\ detectors are very prone to photon pileup when observing bright sources, but when observing with the gratings, photons are dispersed across the entire ACIS detector, often mitigating pileup effects. We estimated the pileup fraction in the first \cha\ epoch (higher flux than the second epoch; Fig. \ref{fig:lt_lc}) to be low, at most only 3\% in the MEG spectrum, at $\lambda\sim3$\,\AA.\footnote{following the methodology shown here: \href{https://cxc.harvard.edu/proposer/threads/binary/index.html}{https://cxc.harvard.edu/proposer/threads/binary/index.html}}

\section{Analysis and Results}
\label{sec:results}

\subsection{Long-term X-ray light curve}

\begin{figure}
    \centering
    \includegraphics[width=0.475\textwidth]{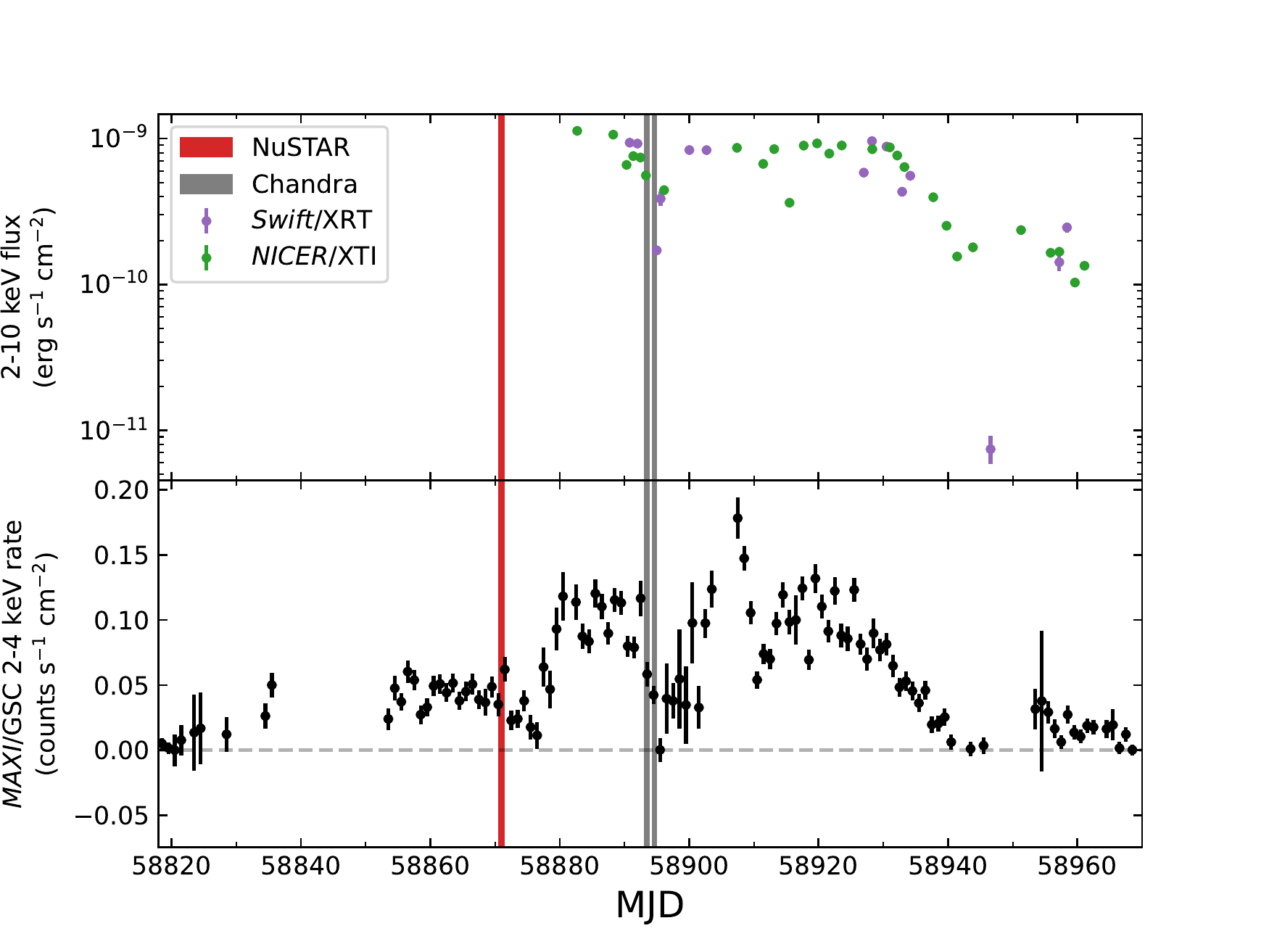}
    \caption{X-ray light curves of \src\ during its 2020 outburst. The upper panel shows the flux of the source in the 2--10\,keV energy range, as measured by \sw\ and \nic. The lower panel shows the 2--4\,keV {\em MAXI}/GSC count rate in 1d bins. In both panels the times of the \nus\ observation and the two \cha/ACIS-HETG observations are shown in red and grey, respectively.}
    \label{fig:lt_lc}
\end{figure}

We compiled an X-ray light curve of the 2020 outburst of \src\ using the available monitoring data from {\em MAXI}/GSC, \sw/XRT and \nic/XTI. For \sw\ and \nic\ we extracted 2--10 keV fluxes by modeling the X-ray spectra. Spectral fitting was performed in {\sc xspec} v12.11.1 \citep{Arnaud-1996} using the $\chi^2$ statistic.

For \sw, all spectra were well fit ($\chi^2$/dof$\sim1$; where dof is degrees of freedom) with a multi-colour disc blackbody model ({\tt diskbb} in {\sc xspec}) with an inner disc temperature range of $kT_{\rm in}\sim1.5-2.0$\,keV across the observations. We used the {\tt tbabs} absorption model \citep{Wilms-2000}, with \citet{Wilms-2000} abundances and \citet{Verner-1996} cross-sections. and fixed the hydrogen column density to $N_{\rm H}=2.5\times10^{21}$\,cm$^{-2}$, consistent with previous observations of the system \citep[e.g.][]{Pahari-2015}. The 2--10\,keV unabsorbed flux was extracted using the {\tt cflux} model and we plot the resultant light curve in Fig. \ref{fig:lt_lc}.

For \nic, we followed a similar fitting methodology, employing $\chi^2$ as the fitting statistic. Spectra were fit with an absorbed disc blackbody model, with a fixed $N_{\rm H}=2.5\times10^{21}$ cm$^{-2}$ and a Gaussian in the 6--7\,keV region to account for the Fe emission seen in this energy range. As with the \sw\ spectral fits, all fits to the \nic\ data were good ($\chi^2$/dof$\sim1$) and $kT_{\rm in}$ was in the range of $\sim1.5-2.0$\,keV. Unabsorbed 2--10\,keV fluxes were again extracted with the {\tt cflux} model. We plot the \nic/XTI light curve alongside the \sw/XRT one in Fig. \ref{fig:lt_lc}.

\subsection{\nus\ spectral analysis}
\label{subsec:NS_analysis}

\begin{figure}
    \centering
    \includegraphics[width=0.475\textwidth]{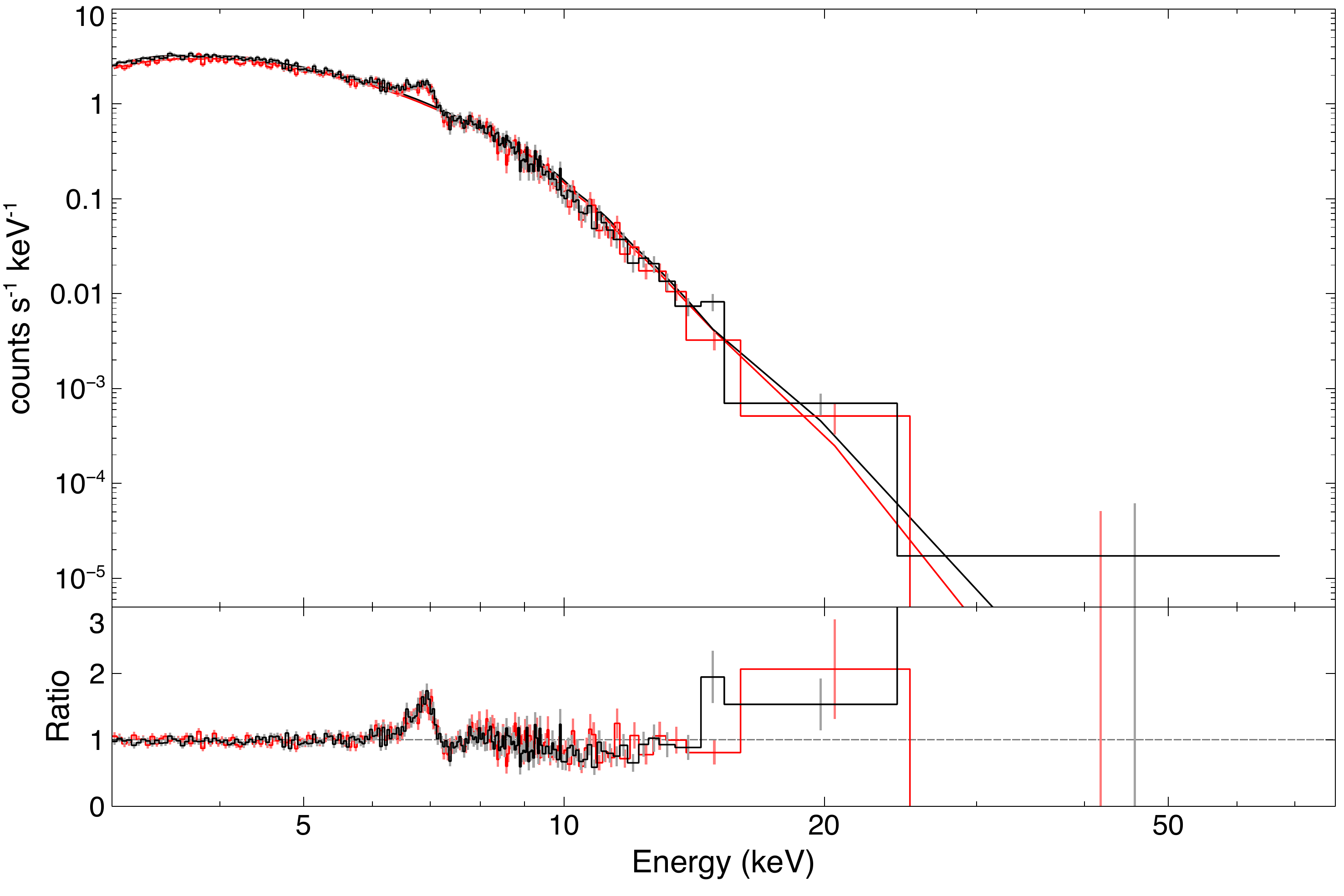}
    \caption{Folded \nus\ FPMA (black) and FPMB (red) spectra of \src. For clarity, only the mode 01 spectra are shown, but models were fit to both mode 01 and mode 06 spectra simultaneously. An absorbed disc blackbody fit to the spectrum is shown in the upper panel as solid lines, and the lower panel shows the data/model ratio for that model.}
    \label{fig:NS_diskbb}
\end{figure}

We present the \nus\ spectra in Fig. \ref{fig:NS_diskbb}, showing only the mode 01 FPMA and FPMB spectra, for clarity. Analysis of the \nus\ spectra was performed with {\sc xspec}, employing $\chi^2$ as the fit statistic. Spectra were not combined, but fit simultaneously, with a cross-normalization factor applied between FPMA and FPMB spectra ({\tt constant} in {\sc xspec}). We attempted to characterize the continuum of the spectrum by fitting an absorbed disc blackbody ({\tt tbabs*diskbb} in {\sc xspec} and {\sc isis} notation). As with the \sw\ and \nic\ spectra from Section \ref{sec:monitoring}, we fix $N_{\rm H}=2.5\times10^{21}$\,cm$^{-2}$ and find a best-fit $kT_{\rm in}=1.39$\,keV. However, the fit is poor ($\chi^2$/dof$=2432.5/1061$), such that we are unable to determine meaningful uncertainties on $kT_{\rm in}$. We show the data/model ratio in the lower panel of Fig. \ref{fig:NS_diskbb}.

From the residuals, we note a number of prominent features. The most striking is the strong emission feature in the $\sim$6--7\,keV range. At \nus\ energy resolution this feature appears to be asymmetric, but we later show that the line can be resolved with \cha\ into 6.7 and  6.97\,keV Fe~{\sc xxv} and Fe~{\sc xxvi} emission lines, respectively, potentially with a small contribution from neutral Fe K$\alpha$. Secondly, we also find evidence of a weak feature at $\sim8$\,keV, similar to the one described by \citet{Pahari-2015} in a 2014 \nus\ observation of \src\ in the soft state.

\begin{table*}
    \centering
    \caption{Best-fit parameters for models fit to \nus\ FPMA and FPMB spectra of \src.}
    \begin{tabular}{l c c c c}
    \hline
    \hline
    Parameter & Model 1 & Model 2 & Model 3 & Model 4 \\
    \hline
    $N_{\rm H}$ ($10^{21}$\,cm$^{-2}$) & $2.5^a$ & $2.5^a$ & $2.5^a$ & $2.5^a$ \\
    $E_{\rm edge}$ (keV) & \ldots & \ldots & $9.55\pm0.19$& $9.55\pm0.19$\\
    $\tau$ & \ldots & \ldots & $0.21^{+0.06}_{-0.05}$ & $0.22\pm0.06$\\
    $kT_{\rm in}$ (keV) & $1.39$ & $1.35\pm0.01$ & $1.37\pm0.01$ & $1.37\pm0.01$\\
    $N_{\rm disc}$ & $17.1$ & $18.8\pm0.4$ & $17.7\pm0.5$ & $18.0^{+0.6}_{-0.5}$\\
    $E_{\rm G,1}$ (keV) & \ldots & $6.83\pm0.02$ & $6.22^{+0.14}_{-0.12}$ & $6.22^{+0.13}_{-0.11}$\\
    $\sigma_{\rm G,1}$ (eV) & \ldots & $150^{+23}_{-25}$ & $100^a$ & $100^a$\\
    $N_{\rm G,1}$ ($10^{-4}$\,photon\,cm$^{-2}$ s$^{-1}$) & \ldots & $14.0\pm0.9$ & $2.9\pm0.8$ & $3.1\pm0.8$\\
    $E_{\rm G,2}$ (keV) & \ldots & $8.10\pm0.06$ & $6.86^{+0.02}_{-0.01}$ & $6.86^{+0.02}_{-0.01}$\\
    $\sigma_{\rm G,2}$ (eV) & \ldots & $<223$ & $100^a$ & $100^a$\\
    $N_{\rm G,2}$ ($10^{-4}$\,photon\,cm$^{-2}$ s$^{-1}$  &  \ldots & $1.9\pm0.5$ & $12.1\pm0.7$ & $12.3^{+0.7}_{-0.8}$\\
    $E_{\rm G,3}$ (keV) & \ldots & \ldots & $8.09\pm0.07$ & $8.09\pm0.06$\\
    $\sigma_{\rm G,3}$ (eV) & \ldots & \ldots & $100^a$ & $100^a$\\
    $N_{\rm G,3}$ ($10^{-4}$\,photon\,cm$^{-2}$ s$^{-1}$ & \ldots & \ldots & $1.4\pm0.4$ & $1.5\pm0.4$ \\
    $\Gamma$ & \ldots & \ldots & \ldots & $1.56^{+1.33}_{-1.39}$ \\
    $N_{\rm pow}$ & \ldots & \ldots & \ldots & $3.3\times10^{-6}$ -- $0.014$\\
    $C_{\rm FPMB}$ & $1.00$ & $1.00\pm0.01$ & $1.00\pm0.01$ & $1.00\pm0.01$\\
    \hline
    $\chi^2$/dof & $2432.5/1061$ & $1337.6/1055$ & $1238.8/1053$ & $1223.3/1051$\\
    \hline\\[-8pt]
    \multicolumn{5}{l}{Model 1: {\tt const*tbabs*diskbb}}
    \\\multicolumn{5}{l}{Model 2: {\tt const*tbabs*(diskbb+gauss+gauss)}}
    \\\multicolumn{5}{l}{Model 3: {\tt const*tbabs*edge*(diskbb+gauss+gauss+gauss)}}
    \\\multicolumn{5}{l}{Model 4: {\tt const*tbabs*edge*(diskbb+gauss+gauss+gauss+powerlaw)}}
    \\\multicolumn{5}{l}{$N_{\rm H}$: Hydrogen column density}
    \\\multicolumn{5}{l}{$kT_{\rm in}$: Inner disc temperature}
    \\\multicolumn{5}{l}{$N_{\rm disc}$: Normalization of the {\tt diskbb} component}
    \\\multicolumn{5}{l}{$E_{\rm G,1/2/3}$: Central energy of Gaussian component 1/2/3}
    \\\multicolumn{5}{l}{$\sigma_{\rm G,1/2/3}$: Width of Gaussian component 1/2/3}
    \\\multicolumn{5}{l}{$N_{\rm G,1/2/3}$: Normalization of Gaussian component 1/2/3}
    \\\multicolumn{5}{l}{$\Gamma$: Power law index}
    \\\multicolumn{5}{l}{$N_{\rm pow}$: Normalization of the power law component}
    \\\multicolumn{5}{l}{$C_{\rm FPMB}$: Cross-normalization factor between FPMA and FPMB}
    \\\multicolumn{5}{l}{$\chi^2/{\rm dof}$: $\chi^2$/degrees of freedom}
    \\\multicolumn{5}{l}{$^{a}$ fixed}
    \end{tabular}
    
    \label{tab:NS_fits}
\end{table*}

To determine the best-fit model for the \nus\ spectrum of \src\ we incrementally add model components to the original absorbed disc blackbody fit shown in Fig. \ref{fig:NS_diskbb}. We first focus on attempting to model the emission features. Including two broad Gaussians in the model, similar to the approach taken by \citet{Pahari-2015}, improves the fit significantly ($\chi^2$/dof$=1337.6/1055$). We find best-fit central energies of $6.83\pm0.02$ and $8.10\pm0.06$\,keV for each Gaussian, with widths of $\sigma=150^{+23}_{-25}$ and $<223$ eV, respectively. \citet{Pahari-2015} suggest that their $8.3$\,keV feature could be due to highly ionized nickel. However, knowing from the \cha/HETG (Section \ref{sec:Fe}) that the $6.83$\,keV feature is largely a blend of highly ionized Fe~{\sc xxv} Ly$\alpha$ and Fe~{\sc xxvi} He$\alpha$, we suggest here that the $8.1$\,keV line that we detect is a blend of Fe~{\sc xxv} He$\beta$ and Fe~{\sc xxvi} Ly$\beta$ at $7.88$ and $8.25$\,keV, respectively. It is possible that there is some contribution from nickel, but \nus's spectral resolution precludes us from resolving individual lines.

We attempt to improve the fit by modeling the Fe 6--7\,keV region as three narrow (i.e. with a fixed width $\sigma=100$ eV) Gaussians and the (likely Fe) 8.1\,keV feature as two equally narrow Gaussians. However, we find that the fit prefers only two Gaussians in the 6--7\,keV range, one at $6.22^{+0.14}_{-0.12}$\,keV and the other at $6.86^{+0.02}_{-0.01}$\,keV. In addition, we are unable to separate the two components that make up the 8.1\,keV feature, and instead the fit prefers a single narrow Gaussian at $8.09\pm0.06$\,keV ($\chi^2$/dof$=1280.9/1055$). We also find that the addition of an absorption edge ({\tt edge} in {\sc xspec}) at $E_{\rm edge}=9.55\pm0.19$ produces a statistically significant improvement to the fit ($\Delta\chi^2=-42.1$ for 2 fewer dof).

Finally, we test the addition of a power law to the model, but find that the improvement to the fit is only marginal (while $\Delta\chi=-15.5$ for 2 fewer dof, the power-law normalization is poorly constrained). We find a best-fit power law index $\Gamma=1.56^{+1.33}_{-1.39}$ and a normalization $N_{\rm pow}$ in the range $3.3\times10^{-6}$--$0.014$ (90\% confidence). It is likely that we would be able to measure the power law with greater confidence with a longer effective exposure time. However, the extremely soft spectrum shown by \src\ in January 2020 combined with only $7.5$ks of effective exposure time means we are unable to strongly constrain the power law tail. We note that the focus of this work remains on the high-resolution \cha/HETG spectra, so we only present the \nus\ data as a broadband overview of the X-ray spectrum of \src. The best-fit parameters for model fits to the \nus\ data are presented in Table \ref{tab:NS_fits}.

\subsection{High-resolution X-ray spectroscopy}
\label{subsec:Chan_analysis}

\begin{figure}
    \centering
    \includegraphics[width=0.475\textwidth]{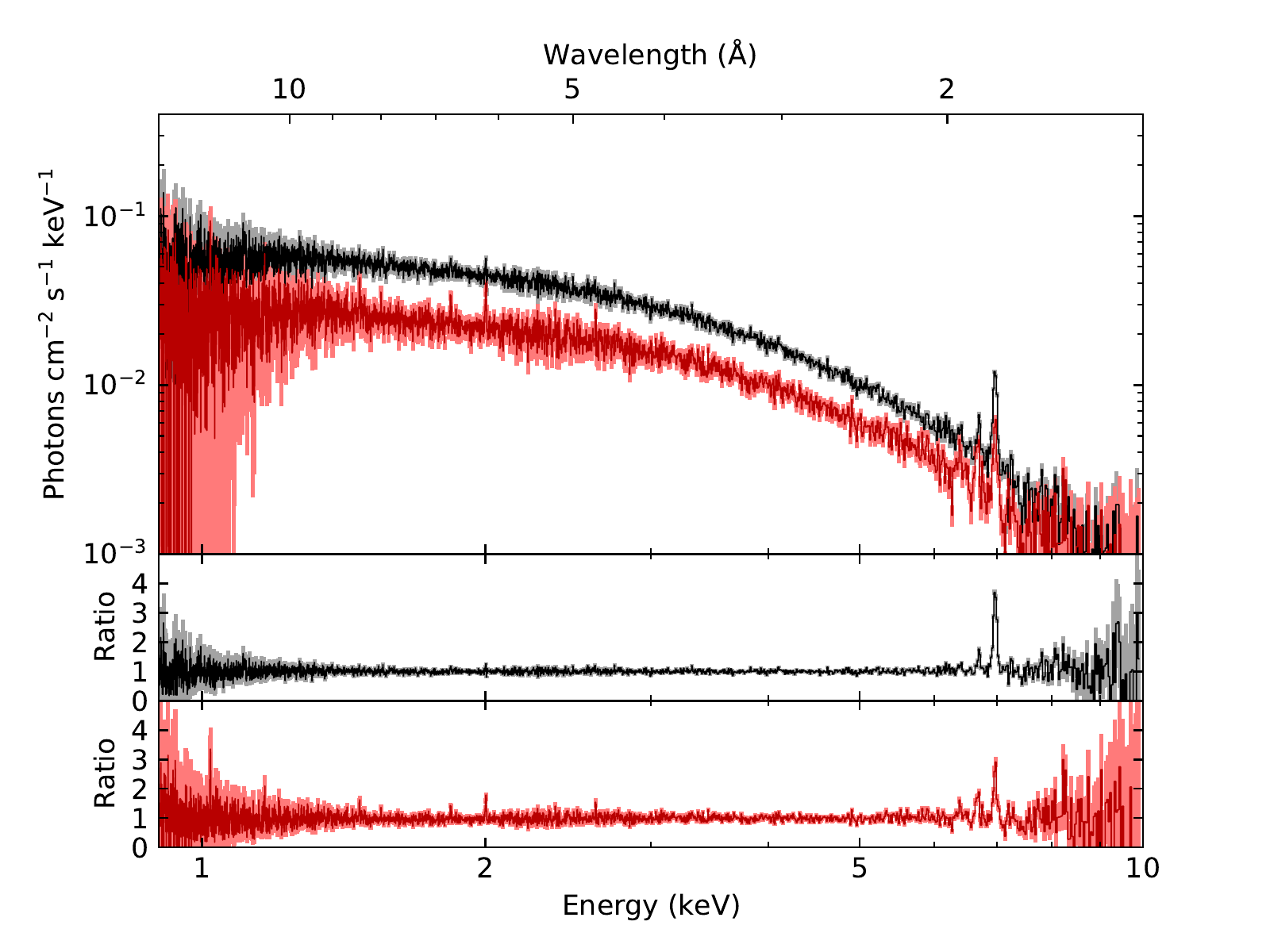}
    \caption{\cha/HETG spectra of \src\ from ObsID 22389 (black) and 23158 (red). The lower two panels show the data/model ratio for a partially covered disc blackbody fit to the continuum for each epoch.}
    \label{fig:full_chandra}
\end{figure}

\cha/HETG spectra were analyzed using using the Interactive Spectral Interpretation System \citep[ISIS;][]{Houck-2000}. We treated each epoch separately, as we know from the \nic\ monitoring that the shape of the X-ray spectrum of \src\ can vary on timescales of less than a day. To extract the most S/N from the high-resolution spectra, we combined the positive and negative first-order HEG and MEG spectra using {\tt combine\_datasets}, after rebinning the HEG spectra on to the MEG wavelength grid using {\tt match\_dataset\_grids}. This practice is well documented in other analyses of \cha/HETG data \citep[see e.g.][]{Miskovicova-2016,Grinberg-2017}.  \cite{Nowak-2019} note that these functions have been well-tested against {\sc HEAsoft} and {\sc ciao} functions that combine spectra and their associated responses and backgrounds. We utilize \citet{Cash-1979} statistics as the fit statistic when modeling the \cha/HETG spectra.

The 0.9--10\,keV \cha\ spectra for each epoch are shown in Fig. \ref{fig:full_chandra}, where a number of strong emission lines are visible. Prior to studying the emission lines we first characterized the X-ray continuum. Similar to the \nus\ spectra, we find that the \cha\ spectra can be well-described by a disc blackbody model, as expected for a BH-LMXB in the soft state. However, we find that we also need to include a partial covering component ({\tt pcfabs}) to fully account for the shape of the continuum in both epochs. Partial covering has been measured in the spectrum of \src\ previously \citep[e.g.][]{Morningstar-2014}, though with a significantly higher covering fraction, ${f=0.97^{+0.01}_{-0.01}}$. Partial covering has also been observed in other X-ray binaries with massive companions \citep[e.g. Vela X-1 and Cyg X-1;][]{Malacaria-2016,Hirsch-2019}. The best-fit continuum model parameters for each epoch are presented in Table \ref{tab:cont_model} and are broadly consistent across the two epochs, aside from the lower normalization of the disc blackbody and a slightly hotter disc in Epoch~1.

\begin{table*}
    \centering
    \caption{Best-fit spectral parameters for the 0.9--10\,keV continuum of \src, for both \cha/HETG epochs. The residuals of this fit are plotted in Figure \ref{fig:full_chandra}.}
    \begin{tabular}{l c c}
         \hline
         \hline
         Parameter & ObsID 22389 & ObsID 23158 \\
          & (Epoch~1) & (Epoch~2) \\
         \hline
         $N_{\rm H}$ ($10^{21}$\,cm$^{-2}$) & $3.1^{+0.1}_{-0.4}$ &  $2.8^{+0.4}_{-0.3}$\\
         $N_{{\rm H}, {\tt pcfabs}}$ ($10^{21}$\,cm$^{-2}$) & $53.9^{+10.7}_{-6.7}$ & $69.4^{+7.0}_{-14.8}$ \\
         $f$ & $0.32\pm0.02$ & $0.37^{+0.04}_{-0.03}$ \\
         $kT_{\rm in}$ (keV) & $1.42^{+0.02}_{-0.01}$ & $1.52\pm0.04$ \\
         $N_{\rm disc}$ & $11.6^{+0.6}_{-0.9}$ & $5.0\pm0.7$ \\
         \hline
         $F_{\rm0.9-10, unabs}$ ($10^{-10}$\,erg\,cm$^{-2}$\,s$^{-1}$) & $8.49^{+0.12}_{-0.16}$& $4.86^{+0.18}_{-0.10}$ \\
         $F_{\rm2-10, unabs}$ ($10^{-10}$\,erg\,cm$^{-2}$\,s$^{-1}$) & $6.07^{+0.09}_{-0.05}$ & $3.62^{+0.07}_{-0.08}$ \\
         $C$/dof & 3357.7/2545& 3427.591/2545\\
         \hline\\[-8pt]
    \multicolumn{3}{l}{$N_{{\rm H}, {\tt pcfabs}}$: Hydrogen column density of the partial covering component}
    \\\multicolumn{3}{l}{$f$: Covering fraction}
    \\\multicolumn{3}{l}{$F_{\rm 0.9-10, unabs}$: Unabsorbed 0.9--10\,keV flux}
    \\\multicolumn{3}{l}{$F_{\rm 2-10, unabs}$: Unabsorbed 2--10\,keV flux}
    \\\multicolumn{3}{l}{$C$/dof: $C$-statistic/degrees of freedom}
    \label{tab:cont_model}
    \end{tabular}
\end{table*}

It is clear from Fig. \ref{fig:full_chandra} that there are emission features present in the spectra from both \cha\ epochs. From a visual inspection of the residuals in Epoch~1 we note two lines likely associated with H-like Fe~{\sc xxvi} (6.97\,keV = 1.78\,\AA) and He-like Fe~{\sc xxv} (6.70\,keV = 1.85\,\AA). In Epoch~1, in addition to ionized iron we see evidence of neutral Fe K$\alpha$ (6.40\,keV = 1.94\,\AA), a narrow line consistent with Si~{\sc xiv} Ly$\alpha$ (2.00\,keV = 6.18\,\AA) and a possible Ne~{\sc x} Ly$\alpha$ line (1.02\,keV = 12.13\,\AA).

To search for and characterize more lines we adopt a statistical approach in the form of a blind line search based on a Bayesian Blocks (BB) algorithm \citep{Scargle-2013}. Though often associated with characterizing variability in time series data, \citet{Young-2007} used a Bayesian Blocks (BB) algorithm on spectral data using the {\sc sitar} package available in {\sc isis}\footnote{\href{https://space.mit.edu/cxc/analysis/SITAR/}{https://space.mit.edu/cxc/analysis/SITAR/}}. The algorithm tests for the presence of lines against a continuum model, in our case the continuum model discussed above and shown in Fig. \ref{fig:full_chandra}. The algorithm groups the data into `blocks' by determining how far each (unbinned) datapoint lies above or below the continuum defined by a parameter $\alpha$, which acts as a significance threshold. The spectrum is considered to have no significant lines if the algorithm returns a single block, and a line is considered to be present between two block change points. The parameter $\alpha$, set by the user, is defined such that the significance of a line is approximately $p\sim\exp(-2\alpha)$ and therefore the probability of a positive detection is roughly $P\sim1-\exp(-2\alpha)$. In this work we define a positive detection to be at least 95\% significant, therefore we require $\alpha>1.5$. The BB line-search method has been discussed and benchmarked against other methods by \citet{Young-2007}, and has been utilized successfully in high-resolution spectroscopic studies of e.g. Vela X-1 \citep{Grinberg-2017} and 4U\,1700$-$37 \citep{Martinez-Chicharro-2021}.

To initiate the blind line search we divide the 1.25--14\,\AA\ ($\sim0.9-10$\,keV) spectrum into five distinct regions, so named for the strongest line(s) usually found in each of the respective wavelength ranges. The regions are as follows: Fe (1.25--2.5\,\AA), S/Ar (2.5--6\,\AA), Si (6--8\,\AA), Mg (8--10\,\AA) and Ne (10--14\,\AA). We then apply the BB algorithm against the best-fit partially covered disc blackbody continuum model, which is frozen in place, and iterate from $\alpha=10$ to $\alpha=1.5$ (in steps of $\Delta\alpha=-0.1$) until a line is found. Once a line is found by the BB algorithm, we fit a Gaussian\footnote{Typically fixing the width to $\sigma=0.003$\,\AA\ ($\sim\frac{1}{3}$ MEG resolution) unless the line is resolved at MEG resolution $FWHM=0.023$\,\AA\ \citep[see e.g.][]{Grinberg-2017,Amato-2021}} and apply the algorithm again, iteratively adding Gaussian components until no more lines are detected above $\sim$95\% significance. For both \cha\ epochs, we describe our results for each spectral region below. Line identifications and their reference wavelengths, $\lambda_0$, are made using {\tt AtomDB} 3.0.9 \citep{Foster-2012} and identified lines for all wavelength regions are presented in Table \ref{tab:lines}.

\begin{table*}
    \centering
    \caption{Spectral lines detected by the Bayesian Blocks algorithm. Quoted rest wavelengths are from the {\tt AtomDB} WebGUIDE interface \citep{Foster-2012}, which uses \citet{Erickson-1977} for H-like ions and \citet{Drake-1988} for He-like ions.}
    \begin{tabular}{lc|cccc|cccc}
    \hline\hline
    & & \multicolumn{4}{c|}{Epoch~1} & \multicolumn{4}{c}{Epoch~2}\\
    \hline
    Line & $\lambda_{0}$ & $\lambda_{\rm obs}$ & $v$ & Flux &  $\alpha$ & $\lambda_{\rm obs}$ & $v$ & Flux & $\alpha$ \\
    & (\AA) & (\AA) & (km s$^{-1}$) & (ph s$^{-1}$\,cm$^{-2}$) & & (\AA) & (km s$^{-1}$) & (ph s$^{-1}$\,cm$^{-2}$) & \\
    \hline
        Fe~{\sc xxvi} Ly$\beta$ & 1.503 & ... & ... & ... & ... & $1.504\pm0.005$ & $200\pm998$ & $1.12^{+0.78}_{-0.60}\times10^{-4}$ & 1.5\\
        Fe~{\sc xxvi} Ly$\alpha$ & 1.780 & $1.780\pm0.001$ & $0\pm169$ & $6.57^{+0.59}_{-0.75}\times10^{-4}$ & $\geq10$ & $1.781^{+0.001}_{-0.002}$ & $169^{+169}_{-337}$ & $2.53^{+0.58}_{-0.53}\times10^{-4}$ & $\geq10$ \\
        Fe~{\sc xxv} He$\alpha$ & 1.855 & $1.852\pm0.002$ & $-485\pm373$ & $1.63^{+0.32}_{-0.49}\times10^{-4}$ & 7.8 & $1.859^{+0.002}_{-0.003}$ & $646^{+323}_{-485}$ & $1.45^{+0.36}_{-0.47}\times10^{-4}$ & 7.7 \\
        \hline
        S~{\sc xvi} Ly$\alpha$ & 4.729 & ... & ... & ... & ... & $4.732^{+0.002}_{-0.003}$ & $190^{+127}_{-190}$ & $7.57^{+4.84}_{-2.50}\times10^{-5}$ & 4.6\\
        \hline
        Si~{\sc xiv} Ly$\alpha$ & 6.182 & $6.193^{+0.001}_{-0.003}$ & $533^{+49}_{-146}$ & $7.64^{+1.45}_{-2.88}\times10^{-5}$ & 6.5 & $6.188^{+0.001}_{-0.002}$ & $291^{+46}_{-97}$ & $1.72^{+0.23}_{-0.27}\times10^{-4}$ & $\geq10$\\
        Si~{\sc xiii} {\it f} & 6.740 & $6.744^{+0.004}_{-0.005}$ & $178^{+178}_{-222}$ & $4.77^{+2.66}_{-1.69}\times10^{-5}$ & 2.6 & $6.742\pm0.002$ & $89\pm89$ & $0.83^{+0.19}_{-0.24}\times10^{-4}$ & $\geq10$ \\
        ? & ... & $7.805^{+0.016}_{-0.017}$ & ... & $7.98^{+3.18}_{-4.88}\times10^{-5}$ & 2.6 & ... & ... & ... & ... \\ 
        ? & ... & $7.959^{+0.003}_{-0.004}$ & ... & $4.99^{+2.21}_{-2.24}\times10^{-5}$ & 2.6 & ... & ... & ... & ... \\
        \hline
        Mg~{\sc xii} Ly$\alpha$ & 8.421 & ... & ... & ... & ... &  $8.428\pm0.002$ & $249\pm71$ & $1.02^{+0.24}_{-0.22}\times10^{-4}$ & $\geq10$ \\
        ? & ... & $9.470\pm0.003$ & ... & $-7.63^{+2.19}_{-3.17}\times10^{-5}$ & 1.7  & ... & ... & ... & ...\\
        \hline
        ? & ... & ... & ... & ... & ... & $10.633^{+0.004}_{-0.005}$ & ... & $1.08^{+0.35}_{-0.47}\times10^{-4}$ & 4.2 \\
        ? & ... & $11.170^{+0.018}_{-0.007}$ & ... & $2.59^{+0.75}_{-1.07}\times10^{-4}$ & 3.0 & ... & ... & ... & ... \\
        Ne~{\sc x} Ly$\alpha$ & 12.134 & ... & ... & ... & ... &  $12.144\pm0.003$ & $247\pm74$ & $2.61^{+0.79}_{-0.69}\times10^{-4}$ & $\geq10$ \\
        ? & ...  & $12.477^{+0.027}_{-0.015}$ & ... & $2.59^{+0.75}_{-1.07}\times10^{-4}$ & 1.5 & ... & ... & ... & ...\\
        \hline
    \end{tabular}
    \label{tab:lines}
\end{table*}

\subsubsection{Fe region}
\label{sec:Fe}

\begin{figure}
    \centering
     \begin{subfigure}[h]{0.475\textwidth}
         \centering
         \includegraphics[width=\textwidth]{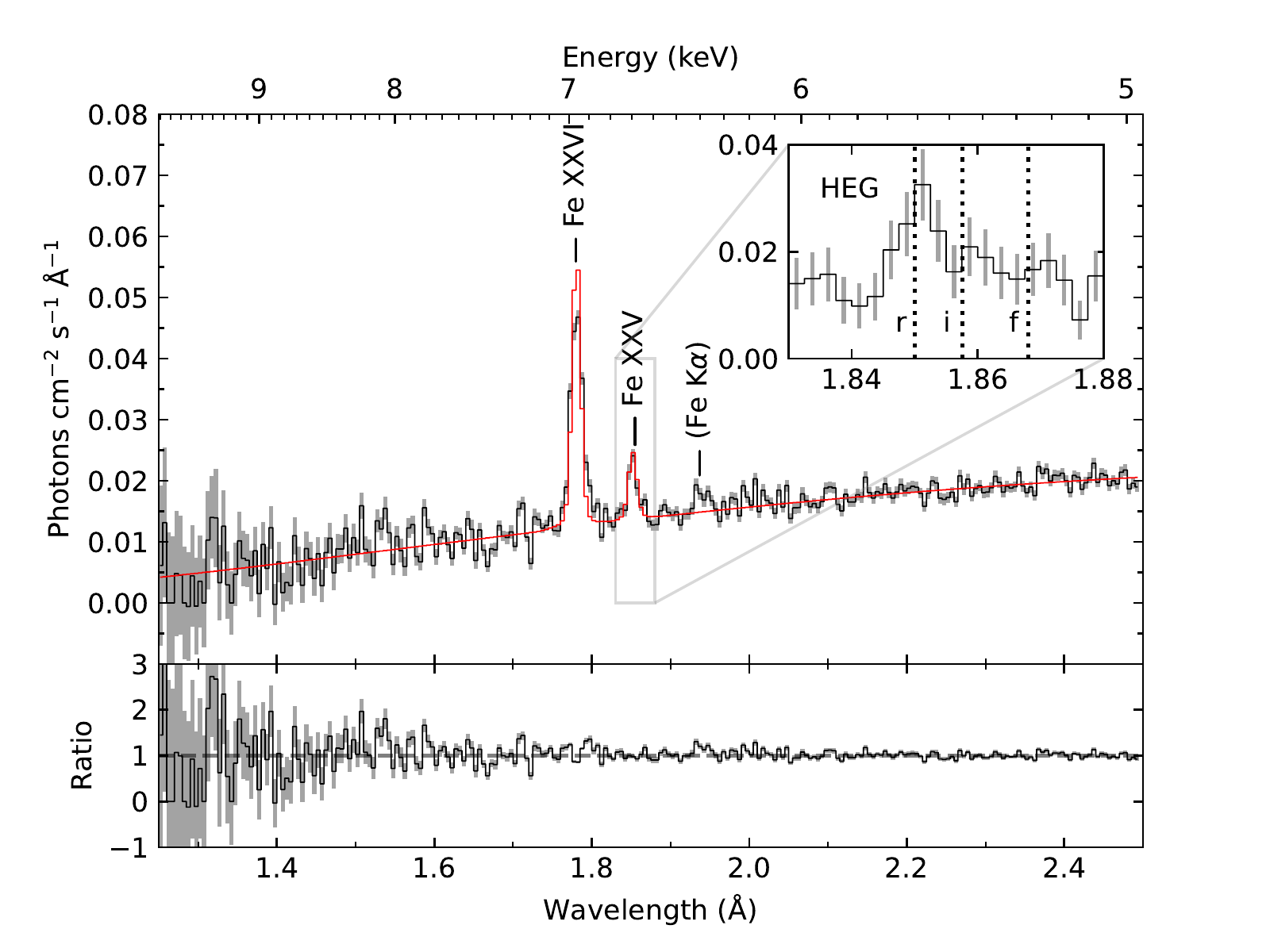}
     \end{subfigure}
     \newline
     \begin{subfigure}[h]{0.475\textwidth}
        \centering
        \includegraphics[width=\textwidth]{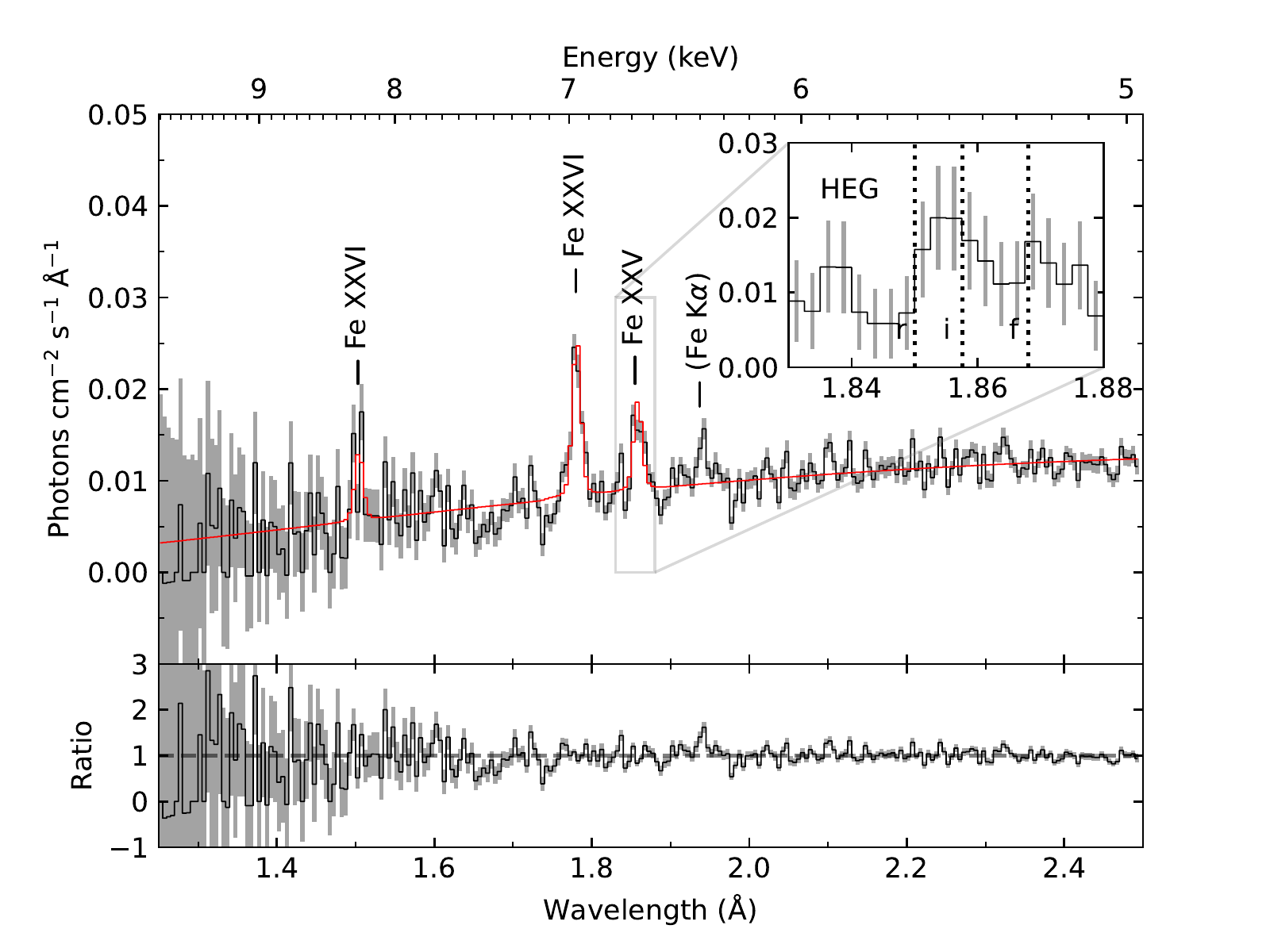}
     \end{subfigure}
     \caption{\cha/HETG spectra of the Fe region. Epoch~1 (ObsID: 22389) is shown in the upper panel and Epoch~2 (ObsID: 23158) in the lower panel. In both panels we show the best-fit model spectrum in red, including lines determined to be significant by the BB algorithm. Features labeled in parentheses are lines that were not detected by the BB algorithm, but could be present in the spectrum given knowledge of expected line energies of particular elements. Data/model ratios are also plotted for each epoch. In each panel, we also plot the HEG-only spectrum of the Fe~{\sc xxv} region inset. Dotted lines show the rest wavelengths of the resonance ($r$), intercombination ($i$) and forbidden ($f$) components of the He-like triplet.}
    \label{fig:Fe_region}
\end{figure}

The spectra of the Fe region are presented in Fig. \ref{fig:Fe_region}. In both epochs, the strongest lines identified are Fe~{\sc xxvi} Ly$\alpha$ and Fe~{\sc xxv} He$\alpha$. Additionally, the BB line search detects a line consistent with Fe~{\sc xxvi} Ly$\beta$ in the second epoch at the lower limit of $\alpha=1.5$. Interestingly, though Fig. \ref{fig:full_chandra} (see also Fig. \ref{fig:Fe_region}) suggests the presence of neutral Fe K$\alpha$, at least in the second epoch, this line is not detected at $\alpha\geq1.5$ by the BB algorithm. Including a Gaussian component at the location of Fe K$\alpha$ results in $\Delta C=-11$ and $\Delta C=-19$ for two fewer degrees of freedom for Epoch~1 and 2, respectively. Regardless of whether Fe K$\alpha$ is present or not, it is clear that the Fe region is dominated by highly ionized iron and  any fluorescent neutral iron is weak.

In both epochs, the H-like ions are consistent with velocity $v=0$~km s$^{-1}$. The He-like Fe~{\sc xxv} lines, conversely, are shifted by ${v=-485\pm373}$ and $v=646^{+323}_{-485}$~km s$^{-1}$ for Epoch~1 and 2, respectively. However, this does not necessarily mean that the lines are indeed Doppler shifted, as He-like Fe~{\sc xxv} consists of a triplet of lines that are not resolved by \cha. The published rest wavelength $\lambda_0=1.855$\,\AA\ is effectively the central wavelength of the triplet and an apparent Doppler shift could simply be explained by a change in relative strength of the forbidden ($\lambda_0=1.868$\,\AA), intercombination ($\lambda_0=1.858$\,\AA), or resonance ($\lambda_0=1.850$\,\AA) components \citep[][]{Drake-1988}. To investigate this possibility,  we examined just the HEG spectra from each epoch, as the HEG has a slightly higher resolution than MEG (0.012\,\AA).\footnote{\href{https://cxc.harvard.edu/proposer/POG/html/chap8.html}{https://cxc.harvard.edu/proposer/POG/html/chap8.html}} The HEG spectra of the Fe~{\sc xxv} region are shown in the insets of both panels of Fig. \ref{fig:Fe_region}, indicating that a change in relative line strengths {\it may} have occurred between the two epochs. However, the spectral resolution of the HEG precludes us from confidently disentangling individual triplet components or performing plasma diagnostics, so we are only able to speculate.

\subsubsection{S/Ar region}

\begin{figure}
    \centering
     \begin{subfigure}[h]{0.475\textwidth}
         \centering
         \includegraphics[width=\textwidth]{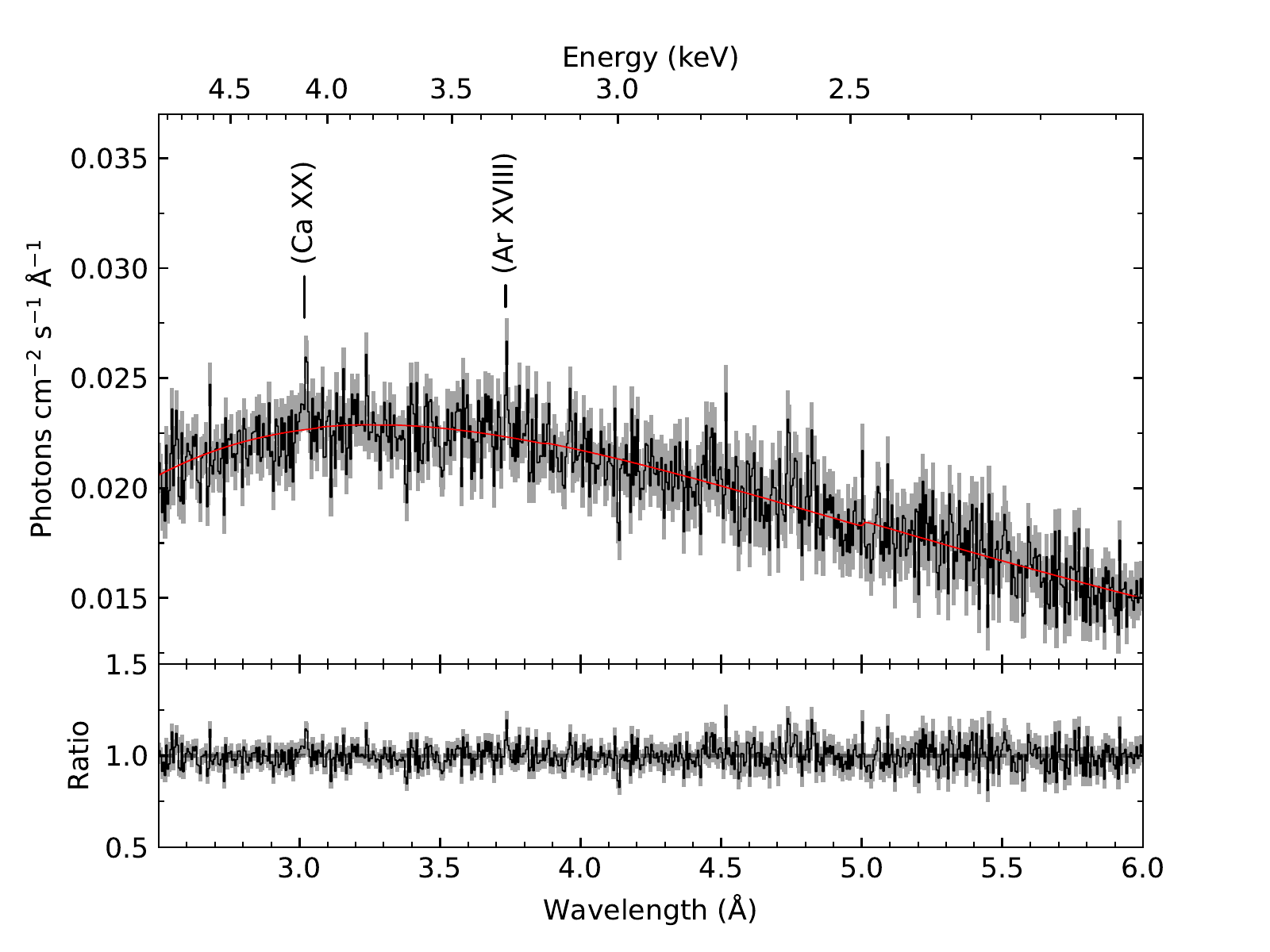}
     \end{subfigure}
     \newline
     \begin{subfigure}[h]{0.475\textwidth}
        \centering
        \includegraphics[width=\textwidth]{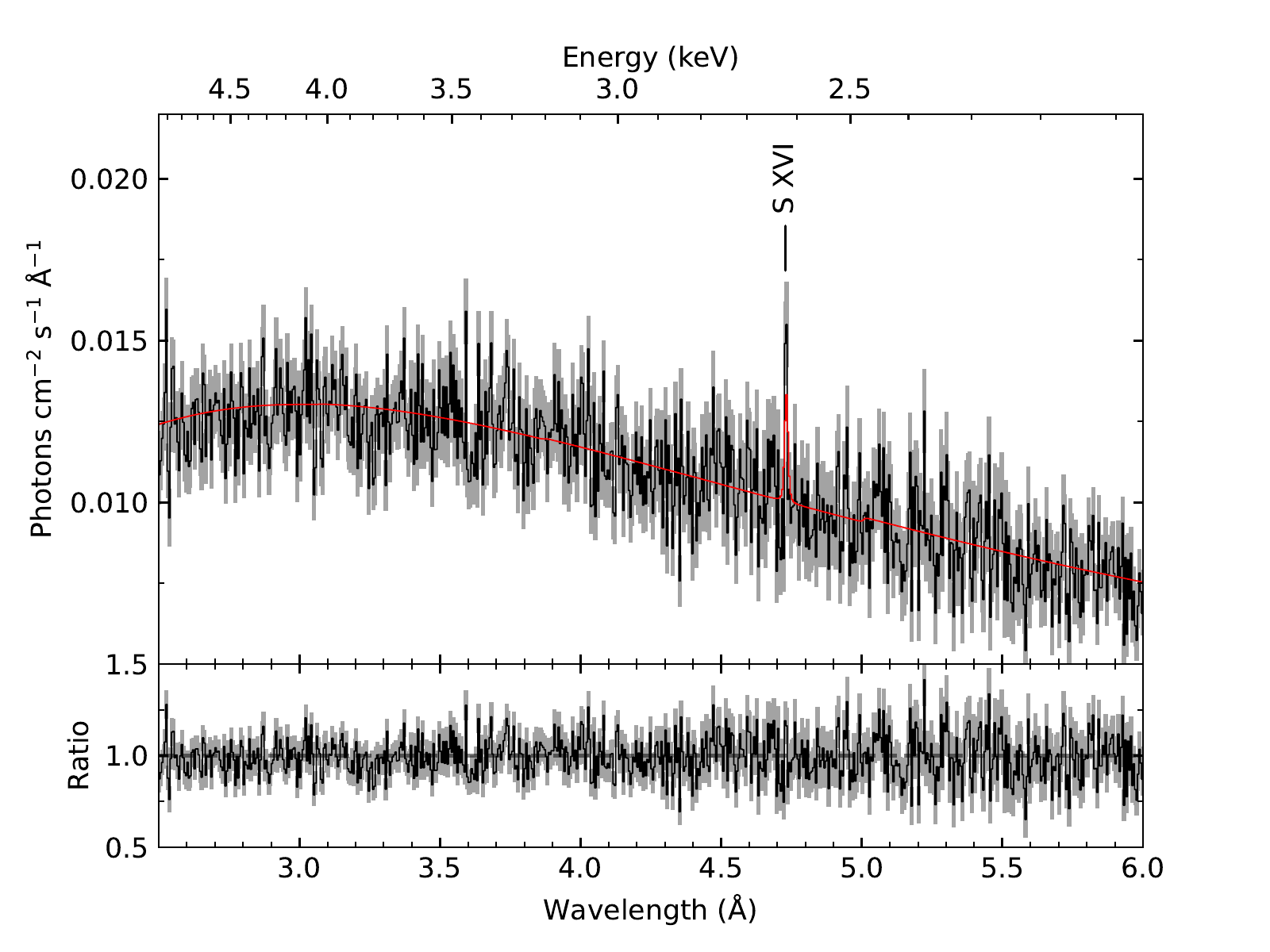}
     \end{subfigure}
     \caption{\cha/HETG spectra of the S/Ar region. Epoch~1 (ObsID: 22389) is shown in the upper panel and Epoch~2 (ObsID: 23158) in the lower panel. In both panels we show the best-fit model spectrum in red, including, where relevant, lines determined to be significant by the BB algorithm. Features labeled in parentheses are lines that were not detected by the BB algorithm, but could be present in the spectrum given knowledge of expected line energies of particular elements. Data/model ratios are also plotted for each epoch.}
    \label{fig:S-Ar_region}
\end{figure}

The BB algorithm does not find statistical evidence for any lines in Epoch~1 ($\alpha<1.5$). However, a visual inspection of the spectra and their residuals (Fig. \ref{fig:S-Ar_region}) suggests that there are lines consistent with the known wavelengths of Ca~{\sc xx} Ly$\alpha$ (3.018\,\AA) and Ar~{\sc xviii} Ly$\alpha$ (3.733 \,\AA) in Epoch~1. If we include two Gaussians in the model for Epoch~1 we find $\Delta C=-27$ for four fewer degrees of freedom ($\Delta C=-14$ and $\Delta C=-13$ for Ca and Ar, respectively), and best-fit velocities of $v=596\pm398$ and $v=321^{+321}_{-241}$\,km\,s$^{-1}$ for Ca~{\sc xx} and Ar~{\sc xviii}, respectively. By Epoch~2, we no longer see evidence for these lines, and instead Ca and Ar have been replaced by a statistically significant S~{\sc xvi} Ly$\alpha$ line with $v=190^{+127}_{-190}$\,km\,s$^{-1}$.

\subsubsection{Si region}
\label{sec:Si}

\begin{figure}
    \centering
     \begin{subfigure}[h]{0.475\textwidth}
         \centering
         \includegraphics[width=\textwidth]{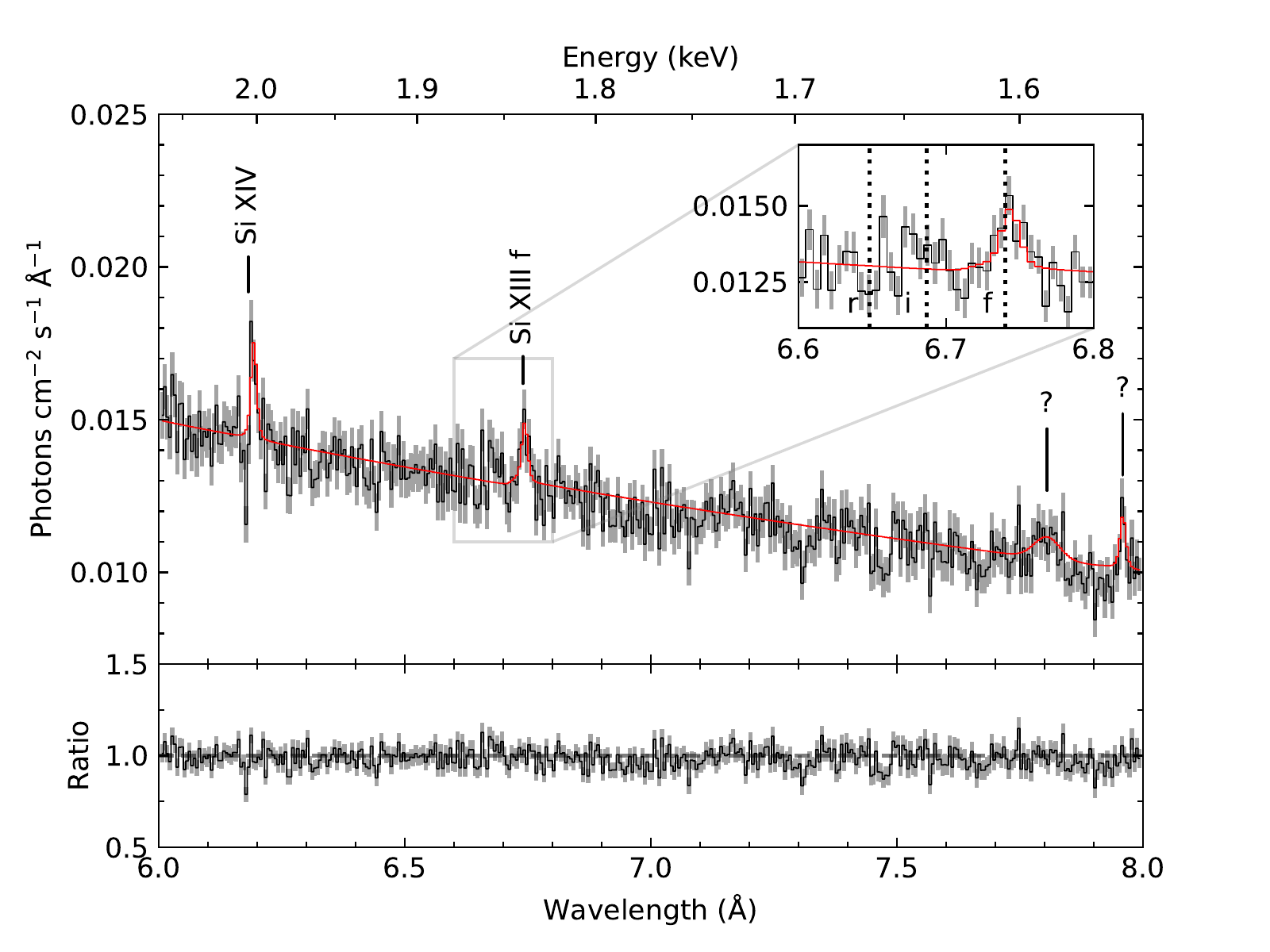}
     \end{subfigure}
     \newline
     \begin{subfigure}[h]{0.475\textwidth}
        \centering
        \includegraphics[width=\textwidth]{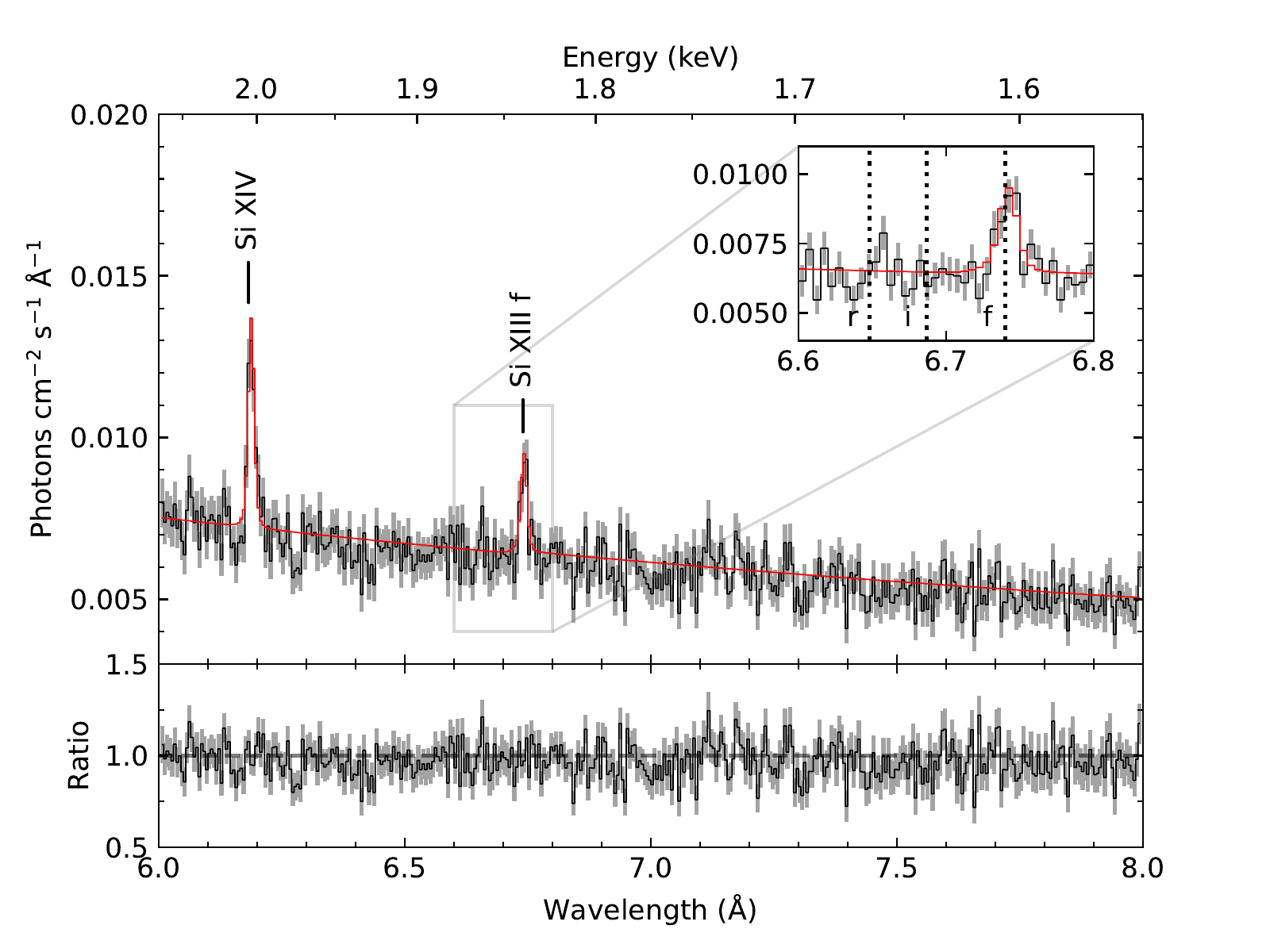}
     \end{subfigure}
     \caption{\cha/HETG spectra of the Si region. Epoch~1 (ObsID: 22389) is shown in the upper panel and Epoch~2 (ObsID: 23158) in the lower panel. In both panels we show the best-fit model spectrum in red, including lines determined to be significant by the BB algorithm. Lines with an uncertain identification are labeled `?' Data/model ratios are also plotted for each epoch. In each panel we show a zoom of the Si {\sc xiii} region inset, with dotted lines showing the rest wavelengths of the resonance ($r$), intercombination ($i$) and forbidden ($f$) components of the He-like triplet.}
    \label{fig:Si_region}
\end{figure}

The Si region exhibits two easily identifiable lines in both epochs (Fig. \ref{fig:Si_region}), those of Si~{\sc xiv} Ly$\alpha$ and the forbidden transition of the He-like Si~{\sc xiii} triplet. In addition, the BB algorithm identifies two features in Epoch~1 that are difficult to reconcile with known features. The first feature is centered at $\lambda=7.805^{+0.016}_{-0.017}$\,\AA\ and is consistent with the He-like Al~{\sc xii} intercombination line. However, there is no evidence for Al~{\sc xiii} Ly$\alpha$ elsewhere in the spectrum and the 7.805\,\AA\ feature is broad in comparison to most of the other lines. Thus, the line is likely not due to Al~{\sc xii}. The second unidentified line is centered at $\lambda=7.959^{+0.003}_{-0.004}$\,\AA. The closest known feature in the {\tt AtomDB} is Li-like Fe~{\sc xxiv}. However, if this was the true identity of the detected line, it would have $v=-1127^{+113}_{-150}$\,km\,s$^{-1}$, very inconsistent with other lines. As such, we consider the two features measured in Epoch~1 to remain unidentified.

The rest wavelength of Si~{\sc xiii} $f$ is almost identical to that of Mg~{\sc xii} Ly$\gamma$. However, if the measured line is indeed Mg~{\sc xii} Ly$\gamma$, then we would also expect the stronger Mg~{\sc xii} Ly$\beta$ at $7.1$\,\AA, which is not seen. The presence of only the forbidden member of the triplet and a negligible contribution from the resonance line ($\lambda_0=6.648$\,\AA) is indicative of a photoionized plasma \citep{Porquet-2000}. We discuss plasma diagnostics using the He-like Si lines in Section \ref{sec:diagnostics}.

\subsubsection{Mg region}

\begin{figure}
    \centering
     \begin{subfigure}[h]{0.475\textwidth}
         \centering
         \includegraphics[width=\textwidth]{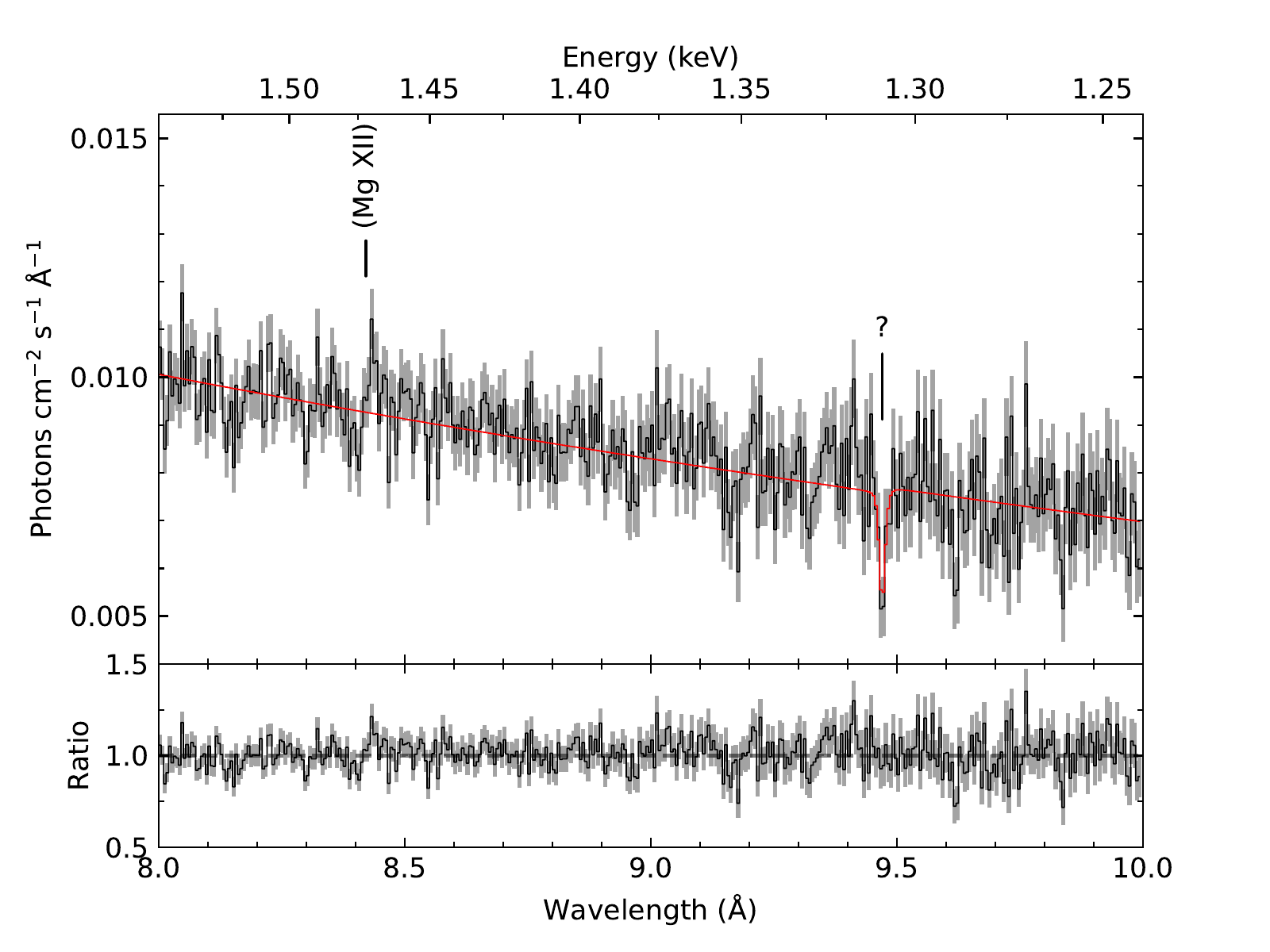}
     \end{subfigure}
     \newline
     \begin{subfigure}[h]{0.475\textwidth}
        \centering
        \includegraphics[width=\textwidth]{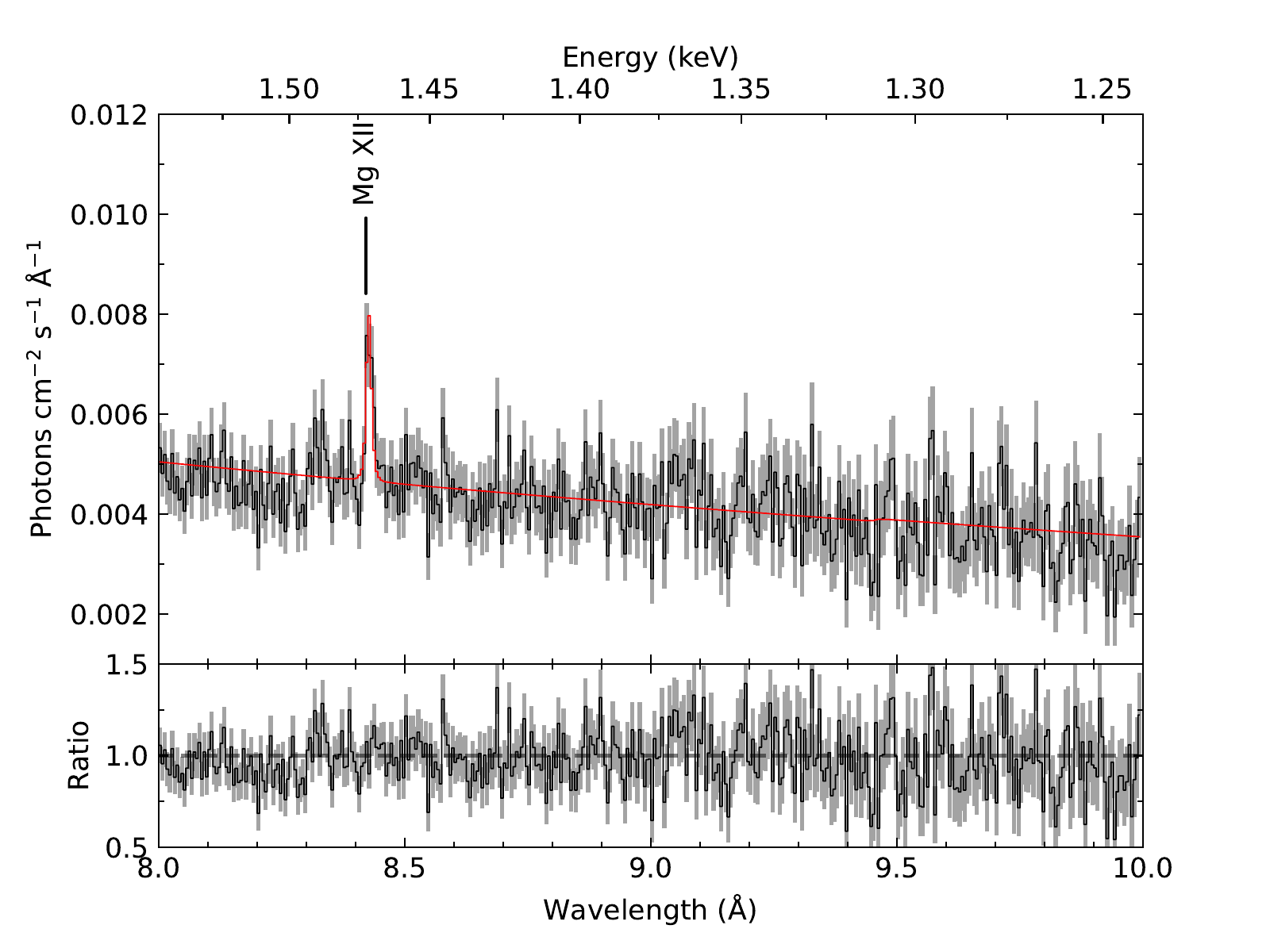}
     \end{subfigure}
     \caption{\cha/HETG spectra of the Mg region. Epoch~1 (ObsID: 22389) is shown in the upper panel and Epoch~2 (ObsID: 23158) in the lower panel. In both panels we show the best-fit model spectrum in red, including lines determined to be significant by the BB algorithm. Features labeled in parentheses are lines that were not detected by the BB algorithm, but could be present in the spectrum given knowledge of expected line energies of particular elements. Lines with an uncertain identification are labeled `?' Data/model ratios are also plotted for each epoch.}
    \label{fig:Mg_region}
\end{figure}

In the Mg region, the BB algorithm identifies two lines (Fig. \ref{fig:Mg_region}). The first, in Epoch~1, is an absorption line at $9.470\pm0.003$\,\AA, which we are unable to identify with any line transitions commonly seen in LMXBs. {\tt AtomDB} does note some $n=5\rightarrow1$ transitions of Ne~{\sc x} at similar energies, but we do not detect any stronger $n=4,3\rightarrow1$ transitions in the spectra. In addition, Ne~{\sc x} Ly $\alpha$ only appears (in emission) in Epoch~2, so the absorption feature remains puzzling.

The second line, detected by the BB algorithm only in Epoch~2, corresponds to Mg~{\sc xii} Ly$\alpha$ with $v=249\pm71$\,km\,s$^{-1}$. Though this feature is not detected by the BB algorithm in Epoch~1 (with a threshold for detection $\alpha\geq1.5$), the residuals in the top panel of Fig. \ref{fig:Mg_region} suggest that a weak line may be present. Including a Gaussian line in the model results in $\Delta C=-14$ for two fewer degrees of freedom, and a best-fit $v=499\pm143$\,km\,s$^{-1}$. This line, if real, is much weaker in Epoch~1 (Flux$=4.81^{+2.35}_{-2.06}\times10^{-5}$ photons s$^{-1}$\,cm$^{-2}$) than in Epoch~2 (Flux$=1.02^{+0.24}_{-0.22}\times10^{-4}$ photons s$^{-1}$\,cm$^{-2}$), something that is also seen in the Si region across epochs.

\subsubsection{Ne region}

\begin{figure}
    \centering
     \begin{subfigure}[h]{0.475\textwidth}
         \centering
         \includegraphics[width=\textwidth]{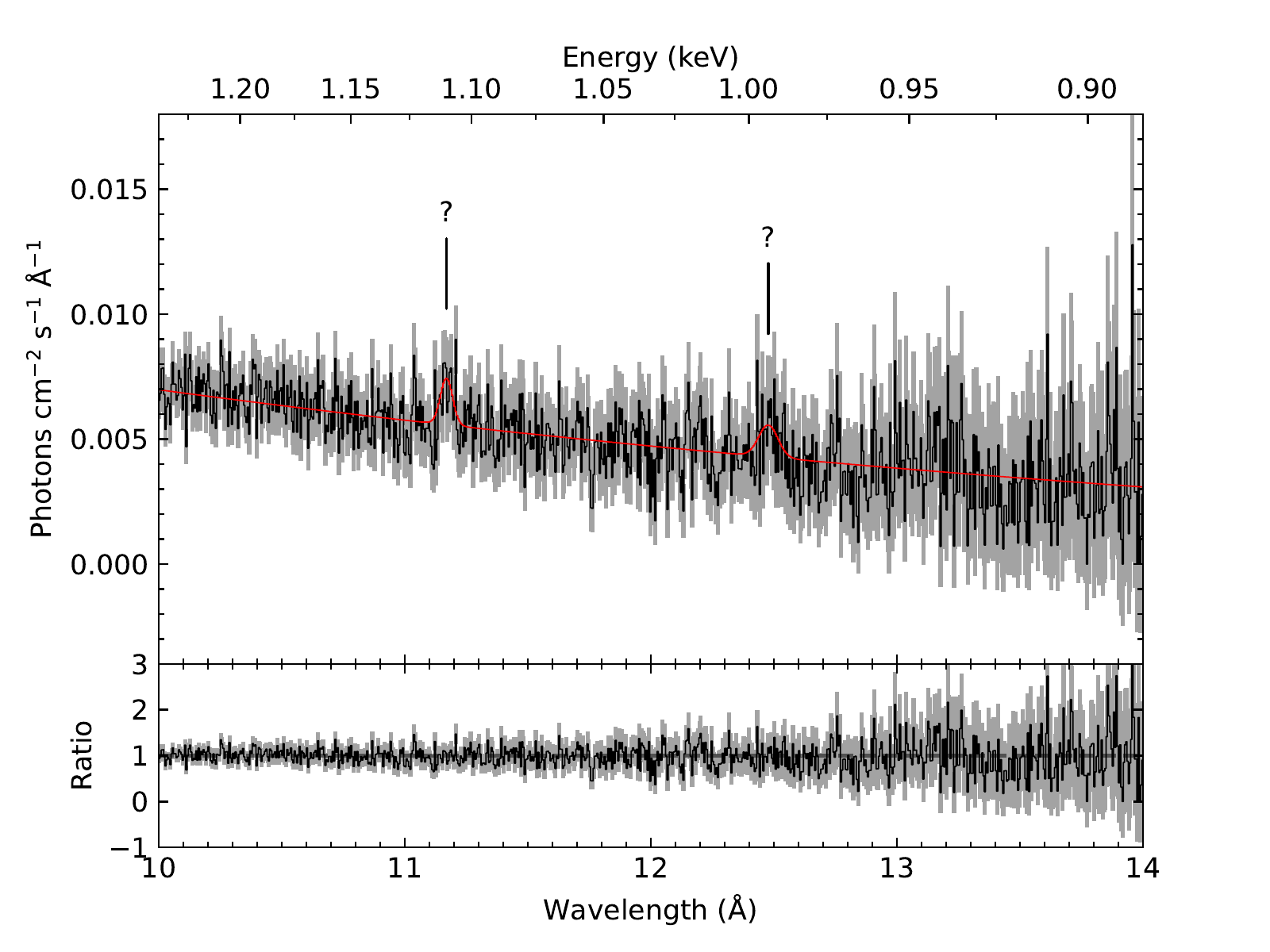}
     \end{subfigure}
     \newline
     \begin{subfigure}[h]{0.475\textwidth}
        \centering
        \includegraphics[width=\textwidth]{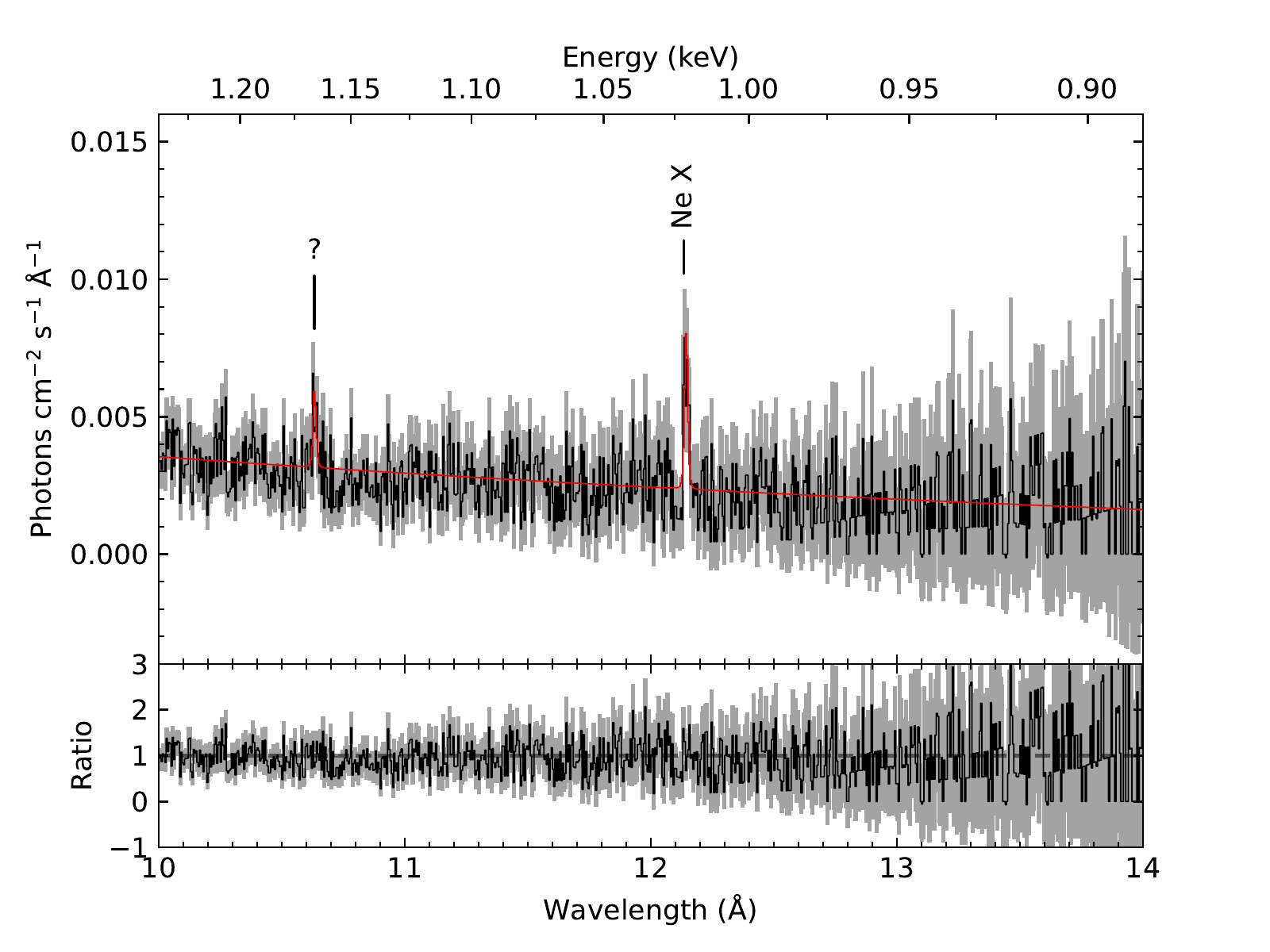}
     \end{subfigure}
     \caption{\cha/HETG spectra of the Ne region. Epoch~1 (ObsID: 22389) is shown in the upper panel and Epoch~2 (ObsID: 23158) in the lower panel. In both panels we show the best-fit model spectrum in red, including lines determined to be significant by the BB algorithm. Lines with an uncertain identification are labeled `?' Data/model ratios are also plotted for each epoch.}
    \label{fig:Ne_region}
\end{figure}

In the Ne region spectra (Fig. \ref{fig:Ne_region}), we can only confidently identify one of the lines detected by the BB algorithm, that of Ne~{\sc x} Ly$\alpha$ at $v=247\pm74$\,km\,s$^{-1}$ in Epoch~2. In Epoch~1, broad lines at $\lambda=11.170^{+0.018}_{-0.007}$ and $12.477^{+0.027}_{-0.015}$\,\AA\ may be associated with the large number of Fe L-shell transitions known to occur around $\sim1$\,keV, but their true nature is unclear. Similarly, the feature at ${\lambda=10.633^{+0.004}_{-0.005}}$\,\AA\ in Epoch~2 is also difficult to identify. At $\lambda\gtrsim12.5$\,\AA\ the source counts are very low, preventing the detection of low-energy features, in particular the Ne~{\sc ix} He-like triplet in the $13.448-13.699$\,\AA\ region.

\section{Discussion}
\label{sec:discussion}

In this section we will discuss the \cha\ spectra and use several diagnostics to determine the nature of the plasma, i.e. whether it is photoionized or if there are other ionization processes at work.

\subsection{A note on the X-ray continuum}
\label{sec:continuum}

In Table \ref{tab:cont_model} we show the best-fit spectral parameters for the \cha\ X-ray continuum indicating that the inner disc temperature in both epochs is $kT_{\rm in}\approx1.4-1.5$. Considering the \citealt{Macdonald-2014} distance to \src\ ($d=6.2\pm0.7$ kpc) and the corresponding $L_{X}\sim4\times10^{35}$\,erg\,s$^{-1}$, the measured $kT_{\rm in}$ corresponds to to a very small inner disc radius $R_{\rm in}\sim 2.9\times10^5$\,cm in Epoch~1 (which is smaller still in Epoch~2). This is smaller than the Schwarzschild radius for a $6.4$ \Msun\ BH ($R_{S}\sim1.9\times10^6$\,cm), and even smaller than the expected innermost stable circular orbit (ISCO) for a maximally rotating Kerr BH ($1.235\,r_{G} \sim1.2\times10^6$\,cm, where $r_{G}=R_{S}/2$). Thus, this $R_{\rm in}$ can be considered unfeasibly small for a source such as \src. We suggest, instead, that the observed flux is coming from X-rays that are scattered above the disc and that the intrinsic X-ray luminosity is actually much higher than measured. This is consistent with other studies of \src\ during the 2020 outburst, namely the in depth analysis of the \nic\ data presented by Buisson et al. (in prep.). In addition, both \citet{Revnivtsev-2002} and \citet{Koljonen-2020} also suggest that the central engine in \src\ may be obscured, potentially by an optically thick equatorial outflow, resulting in the low X-ray luminosities often seen in the system during outburst. We estimate that if the inner disc is at $kT_{\rm in}=1.235r_{G}$ with $kT_{\rm in}=1.4$\,keV, then the intrinsic $L_{X}$ needs to be at least 20 times larger than the value we measure in Epoch~1. In Epoch~2 we estimate that the intrinsic $L_{X}$ should be $>40$ times the measured $L_{X}$. These estimates are important when considering the locations of the line-emitting mediums in Section \ref{sec:phot_modeling}.

\subsection{Plasma diagnostics with He-like triplets}
\label{sec:diagnostics}

As discussed in Section \ref{sec:Si}, resolving He-like lines in X-ray spectra of astrophysical plasmas allows us to determine key properties of the medium, including its density, temperature and the dominant ionization processes \citep{Porquet-2000}. In the case of \src\ we detect a strong line at $\lambda=6.74$\,\AA, consistent with the forbidden, {\it f}, transition of Si~{\sc xiii}. This line is strong in both \cha\ epochs. 

Most notable about the Si region is the apparent lack of emission from the other members of the Si~{\sc xiii} triplet, namely the intercombination ({\it i}; $\lambda_0=6.687$\,\AA) and resonance ({\it r}; $\lambda_0=6.648$\,\AA) transitions. A weak {\it r} line relative to {\it f} and {\it i} is a strong indicator of a purely photoionized plasma. To test this, we utilized the {\tt triplet\_fit} function available as an add-on to ISIS\footnote{\href{https://space.mit.edu/~dph/triplet/triplet.html}{https://space.mit.edu/$\sim$dph/triplet/triplet.html}}, which allows us to measure line ratios directly.

We used a power law to approximate the continuum in the Si region and placed three Gaussian emission lines with $\sigma=0.003$\,\AA\ at the wavelengths of the {\it i}, {\it r} and {\it f}, but shifted by the measured values of $v$ for Si~{\sc xiii} {\it f} as detailed in Table \ref{tab:lines}. We allowed the line flux to vary but fixed $\lambda_{\rm obs}$. In both epochs we find that the best fit line flux for the {\it r} is consistent with zero. A weak resonance line means that collisional processes are negligible and the plasma is likely photoionized \citep{Porquet-2000}. We can quantify this by measuring the ratio ${\rm G}=\frac{{\rm\it f} + {\rm \it i}}{{\rm \it r}}$. A value of G$>$4 is the signature of a purely photoionized plasma. In Epochs~1 and 2 we measure ${{\rm G}>4.37}$ and ${{\rm G}>2.67}$, respectively which, though not definitive in the case of Epoch~2, is evidence for photoionization being the dominant ionization process in the medium. For a purely photoionized plasma, fig.~10 of \citet{Porquet-2000} tells us the expected range of electron temperature, $T_{e}$ and in the case of Si~{\sc xiii} this range is $T_{e}\sim5\times10^5-2\times10^6$ K.

Next we measure the ratio ${\rm R}=\frac{{\rm\it f}}{{\rm \it i}}$, which is an indicator of the particle density of the plasma ($n_{e}$). The strong forbidden lines (relative to other members of the triplet) present in the spectra at both epochs are indicative of a low density plasma. In Epoch~1 we find ${\rm R}=1.97^{+5.32}_{-0.64}$, effectively ${\rm R}>1.35$. According to \citet{Porquet-2000}, this suggests an upper limit of $n_{e}\lesssim5\times10^{13}$\,cm$^{-3}$. In Epoch~2, the {\it i} measured line flux is consistent with zero, meaning we measure a best-fit ${\rm R}>5.33$. This is outside the range at which R is sensitive to density and thus implies a plasma density $n_{e}<10^{12}$\,cm$^{-3}$.

\subsection{Photoionization modeling}
\label{sec:phot_modeling}

\begin{table*}
    \centering
    \caption{Best-fit spectral parameters for the fully modeled 0.9--10\,keV \cha/HETG spectra of \src. The model consists of the partially covered disc-blackbody continuum, plus two photoionized plasmas, the first of which models the non-Fe plasma and includes both an absorption and emission component. The second photoionized plasma models the Fe emission only.}
    \begin{tabular}{l c c}
         \hline
         \hline
         Parameter & Epoch~1 & ObsID Epoch~2 \\
         \hline
         $N_{\rm H}$ ($10^{21}$\,cm$^{-2}$) & $2.8^{+0.0}_{-0.1}$ &  $3.0\pm0.1$\\
         $N_{{\rm H}, {\tt pcfabs}}$ ($10^{21}$\,cm$^{-2}$) & $51.3^{+0.0}_{-1.5}$ & $70.7\pm2.7$ \\
         $f$ & $0.324^{+0.007}_{-0.003}$ & $0.370^{+0.003}_{-0.007}$ \\
         $kT_{\rm in}$ (keV) & $1.441\pm0.001$ & $1.533^{+0.002}_{-0.003}$ \\
         $N_{\rm disc}$ & $10.70^{+0.03}_{-0.10}$ & $54.91^{+0.02}_{-0.09}$ \\
         $z_{\rm abs,1}$ ($\times10^{-4}$)& $-5^{+2}_{-4}$ & $-16^{+8}_{-3}$ \\
         $n_{\rm e,1}$ ($10^{12}$\,cm$^{-3}$) & $6.7^{+0.8}_{-1.7}$ & $5.1^{+1.2}_{-1.1}$ \\
         $N_{\rm e,1}$ ($10^{21}$\,cm$^{-2}$) & $>2.6$ & $>2.8$ \\
         $\log_{10} \xi_{1}$ & $<3.06$ & $<3.14$\\
         $z_{\rm emis,1}$ ($\times10^{-4}$) & $17^{+7}_{-6}$ & $8^{+3}_{-2}$\\
         $N_{\rm XSTAR,1}$ ($\times10^{-3}$) & $5.8^{+1.4}_{-2.2}$ & $12.4^{+1.2}_{-1.6}$ \\
         $n_{\rm e,2}$ ($10^{12}$\,cm$^{-3}$) & $>92.2$ & $7.0^{+0.9}_{-0.8}$ \\
         $N_{\rm e,2}$ ($10^{21}$\,cm$^{-2}$) & $1.3^{+0.2}_{-0.9}$ & $<0.3$ \\
         $\log_{10} \xi_{2}$ & $5.35\pm0.01$ & $5.15^{+0.03}_{-0.05}$\\
         $z_{\rm emis,2}$ ($\times10^{-4}$) & $7^{+4}_{-3}$ & $12^{+2}_{-7}$\\
         $N_{\rm XSTAR,2}$ & $12.9^{+0.0}_{-5.4}$ & $4.9^{+0.6}_{-0.7}$ \\
         \hline
         $C$/dof & 2582.2/2534& 2828.6/2534\\
         \hline\\[-8pt]
    \multicolumn{3}{l}{$z_{{\rm abs},1}$: redshift of the absorption component of the first photoionized plasma}
    \\\multicolumn{3}{l}{$n_{\rm e,1/2}$: electron density of the first/second photoionized plasma}
    \\\multicolumn{3}{l}{$N_{\rm e,1/2}$: column density of the first/second photoionized plasma}
    \\\multicolumn{3}{l}{$\log_{10} \xi_{1/2}$: Logarithm of the ionization parameter for the first/second photoionized plasma}
    \\\multicolumn{3}{l}{$z_{{\rm emis},1/2}$: redshift of the emission component of the first/second photoionized plasma}
    \\\multicolumn{3}{l}{$N_{\rm XSTAR,1/2}$: Normalisation of the first/second photoionized plasma}
    \label{tab:xstar_model}
    \end{tabular}
\end{table*}

Analysis of the He-like Si~{\sc xiii} lines implies that photoionization dominates, at least in the Si-emitting region of the plasma. The R ratio provides an upper limit to the density of the plasma, and the derived G ratio implies $T_{e}\sim5\times10^5-2\times10^6$ K. We can use this information to simulate a photoionized medium and apply it to the spectra of V4641\,Sgr. 

We used the best-fit continuum model as an input to XSTAR v2.54a\footnote{\href{https://heasarc.gsfc.nasa.gov/lheasoft/xstar/xstar.html}{https://heasarc.gsfc.nasa.gov/lheasoft/xstar/xstar.html}} to calculate grids of photoionized spectra. We used the resultant grids to determine the best-fit $n_{e}$ and ionization parameter $\xi$, which is defined according to \citet{Tarter-1969a}:

\begin{equation}
    \xi=\frac{L_{X}}{n_{e} r^2},
    \label{eq:xi}
\end{equation}

\noindent where $r$ is the distance of the line-emitting medium from the central ionizing source.

\begin{figure}
    \centering
     \begin{subfigure}[h]{0.475\textwidth}
         \centering
         \includegraphics[width=\textwidth]{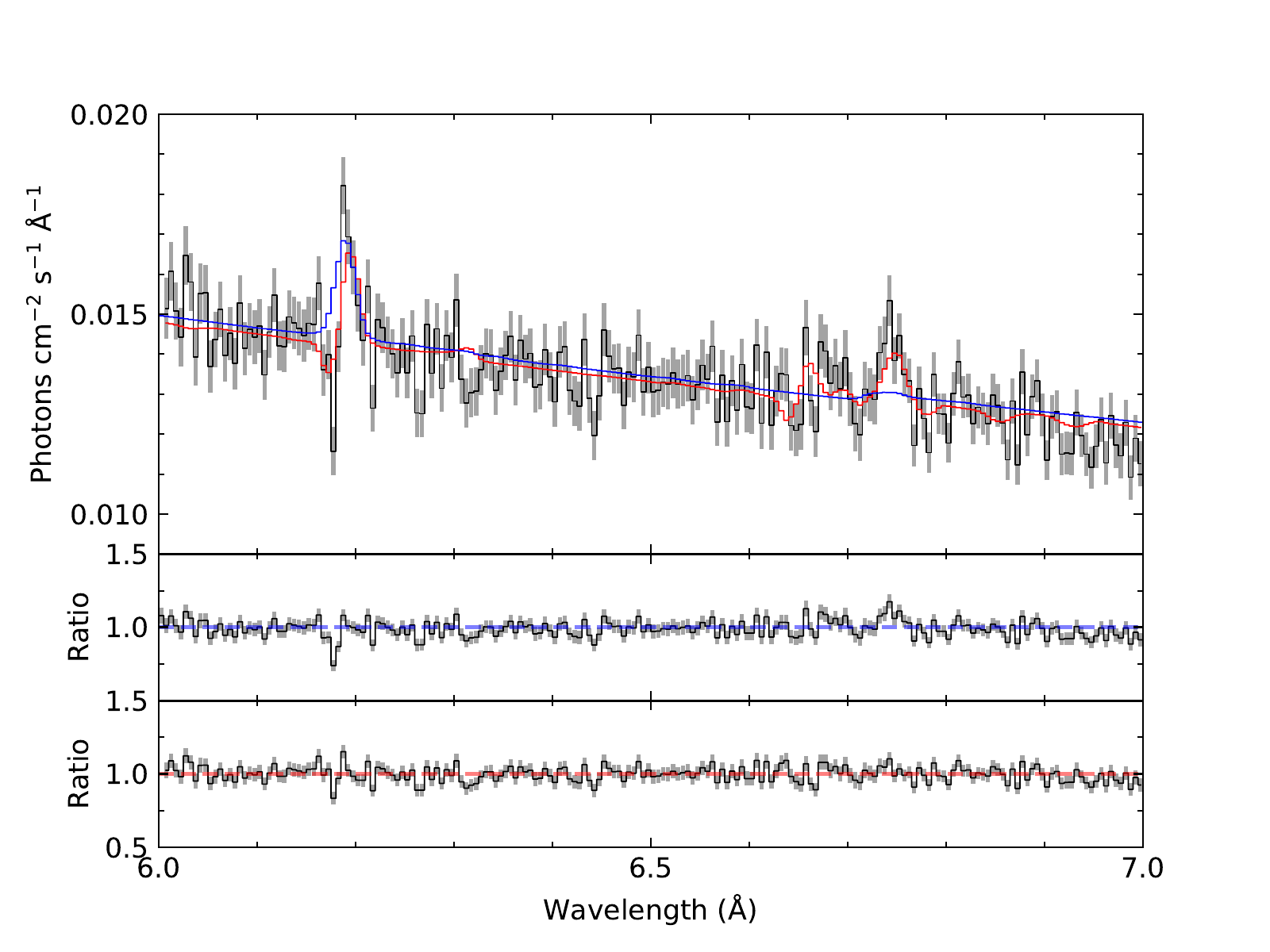}
     \end{subfigure}
     \newline
     \begin{subfigure}[h]{0.475\textwidth}
        \centering
        \includegraphics[width=\textwidth]{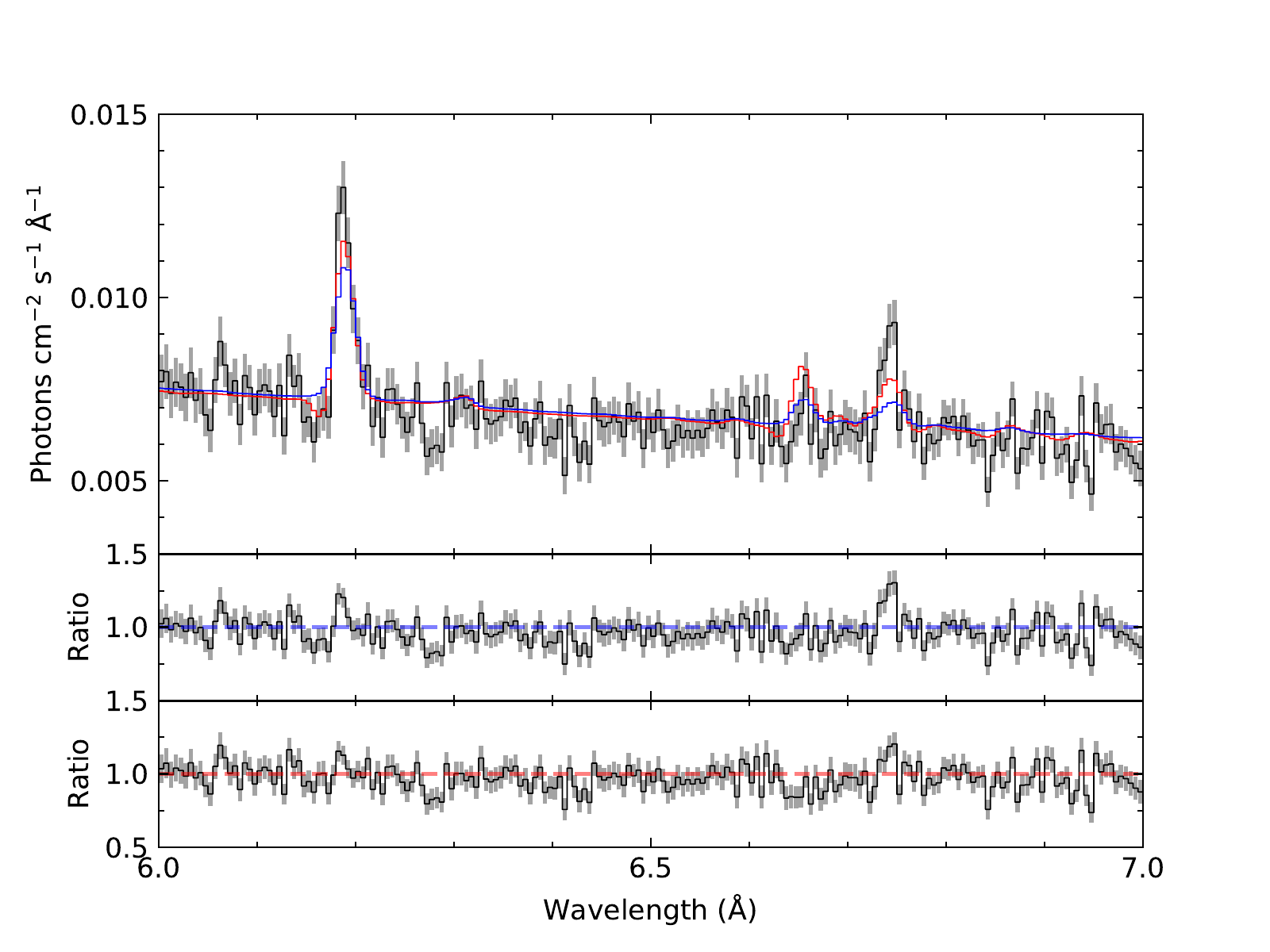}
     \end{subfigure}
     \caption{\cha/HETG spectra of the Si complex. Epochs~1 (ObsID: 22389) and 2 (ObsID: 23158) are shown in the upper and lower panel, respectively. In both panels we show the best-fit {\sc xstar} model found when fitting (a) the full spectrum (blue) and (b) to the Si complex only (red). Data/model ratios are also plotted for each epoch.}
    \label{fig:Si_xstar}
\end{figure}

In Epoch~1, we find that a single photoionized plasma provides an acceptable fit ($C/{\rm dof}=2706/2544$) to the spectrum with ${\log_{10}\xi=5.13^{+0.04}_{-0.01}}$ and ${n_{e}\geq9.5\times10^{13}}$\,cm$^{-3}$. However, this single model is unable to account for the Si~{\sc xiii} He-like lines (Fig. \ref{fig:Si_xstar}, upper panel, blue), which are indicative of a lower density than that implied by the best-fit {\sc xstar} model. Similarly, in Epoch~2, we find $C/{\rm dof}=3015/2544$ for a single photoionized plasma with $\log_{10}\xi=4.80^{+0.16}_{-0.11}$ and $n_{e}=8.9^{+0.2}_{-0.1}\times10^{12}$\,cm$^{-3}$, again find that the Si~{\sc xiii} He-like lines remain poorly constrained in such a model (Fig. \ref{fig:Si_xstar}, lower panel, blue). A single photoionized model applied to the full energy range instead preferentially models the strong Fe lines, in both epochs.

In an attempt to solve this, we consider the possibility that there are multiple emission zones responsible for the lines observed in the spectrum of \src. We therefore apply the photoionized plasma grid generated by {\sc xstar} to just the Si lines, limiting the fit to the \mbox{6--7\,\AA} range. In Epoch~1, we derive a best-fit $n_{e}=6.2^{+0.6}_{-2.8}\times10^{12}$\,cm$^{-3}$, which is consistent with the density derived from R. However, we also find $\log_{10} \xi < 3.18$, much lower than the value derived from fitting to the full spectrum (Fig. \ref{fig:Si_xstar}, upper panel, red). This model does not reproduce the Fe lines when applied to the full spectrum. In the case of Epoch~2, we find similar values for the best-fit {\sc xstar} grid, with $n_{e}=6.2^{+1.4}_{-0.8}\times10^{12}$\,cm$^{-3}$ and $\log_{10} \xi < 3.10$. This is at odds with the $n_{e}<10^{12}$\,cm$^{-3}$ implied by the measured R. However, we note here that (a) the best-fit {\sc xstar} model slightly underestimates the strength of the Si~{\sc xiii} $f$ line (Fig. \ref{fig:Si_xstar}, lower panel, red) and (b) our measured ${\rm R}>5.33$ is higher than any value of R considered by \citet{Porquet-2000} for the Si~{\sc xiii} triplet. Thus, our estimate of $n_{e}$ from line ratios may be inaccurate. Again, the model in Epoch~2 does not reproduce the Fe lines when applied to the full spectrum.

Interestingly, when the model derived just from the Si complex is applied to the full spectrum, we find that, even though the Fe complex is not reproduced, many other lines are. Most notably, the model reproduces the Mg~{\sc xii} Ly$\alpha$ lines detected in both epochs, but we also find that it predicts the S~{\sc xvi} and Ne~{\sc x} lines that are detected by the BB algorithm in Epoch~2 \footnote{However, the unidentified lines detailed in Table \ref{tab:lines} are not reproduced.}. It is therefore possible that multiple emission zones exist within the \src\ system, where the Fe emission is separate to emission from many other metal ions.

\begin{figure*}
    \centering
     \begin{subfigure}[h]{\textwidth}
         \centering
         \includegraphics[width=\textwidth]{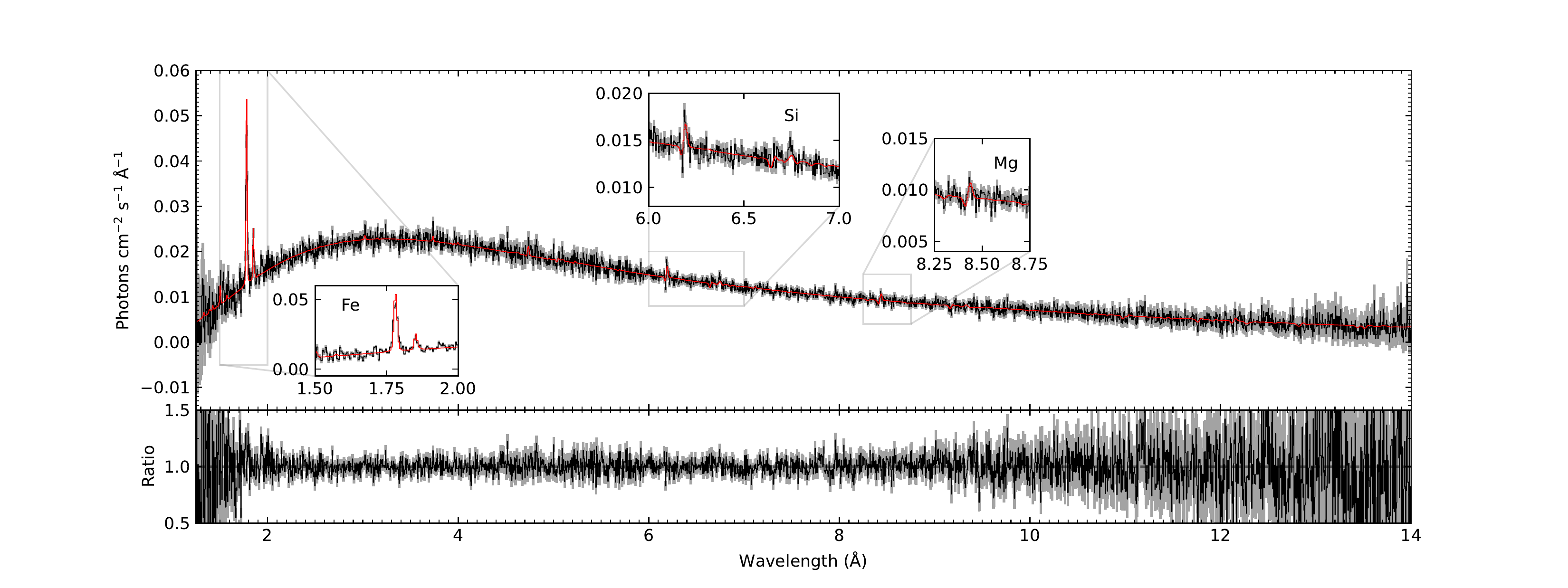}
     \end{subfigure}
     \newline
     \begin{subfigure}[h]{\textwidth}
        \centering
        \includegraphics[width=\textwidth]{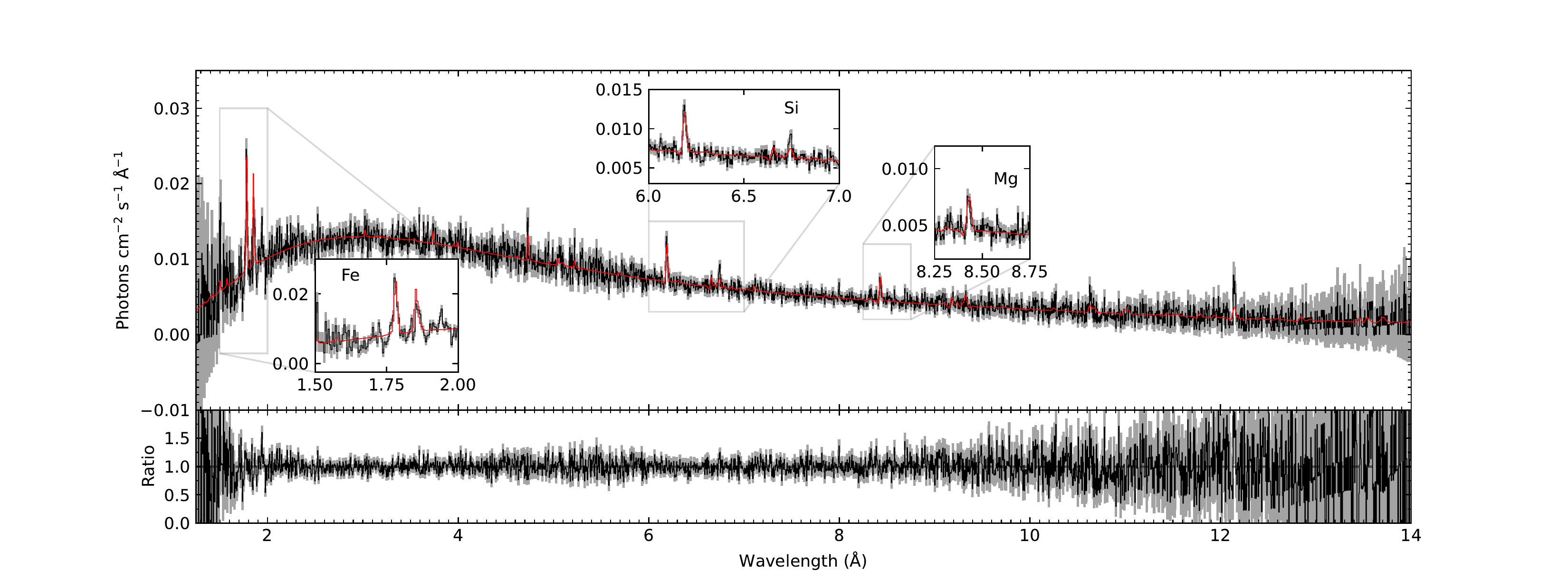}
     \end{subfigure}
     \caption{ Full \cha/HETG spectra of \src. Epochs~1 (ObsID: 22389) and 2 (ObsID: 23158) are shown in the upper and lower panel, respectively. In both panels we show, in red, the best-fit model (presented in Table \ref{tab:xstar_model}) of a partially covered disc blackbody, with two {\sc xstar} photoionized plasmas. In the insets, we show zoom-ins of some of the emission features that have been modeled. Data/model ratios are also plotted for each epoch.}
    \label{fig:full_xstar}
\end{figure*}

We add a second {\sc xstar} component to the model derived from the Si region alone, in an effort to constrain the Fe lines at the same time as those from other ions are constrained. The results of the fit are shown in Table \ref{tab:xstar_model}. We find acceptable fits for both epochs, with $C/{\rm dof}=2582.2/2534$ in Epoch~1 and $C/{\rm dof}=2828.6/2534$ in Epoch~2. In both epochs, the best-fit model implies that there can be two separate photoionized plasma components, each with their own density and ionization parameter. In Epoch~1, we find $n_{\rm e,1}=6.7^{+0.8}_{-1.7}\times10^{12}$\,cm$^{-3}$ and $\log_{10} \xi_{1} < 3.06$ for the first photoionized component, which we interpret as the Si-emitting plasma. For the second photoionized model component, we find $n_{\rm e,2}>9.2\times10^{13}$\,cm$^{-3}$ and $\log_{10} \xi_{2} = 5.35\pm0.01$, which we interpret as the Fe-emitting plasma. For Epoch~2 we find $n_{\rm e,1}=5.1^{+1.2}_{-1.1}\times10^{12}$\,cm$^{-3}$ and $\log_{10} \xi_{1} < 3.14$ for the first (Si-emitting) plasma component and ${n_{\rm e,2}=7.0^{+0.9}_{-0.8}\times10^{12}}$\,cm$^{-3}$ and ${\log_{10} \xi_{2} = 5.15^{+0.03}_{-0.05}}$ for the second (Fe-emitting) plasma component. We show the full fit to the \cha/HETG spectra from both epochs in Fig. \ref{fig:full_xstar}.

Utilizing Equation \ref{eq:xi}, we can estimate the distance, $r$, of the presumed two line emitting mediums from the central ionizing source. Recalling, from Section \ref{sec:continuum}, that the intrinsic $L_{X}$ is likely more than an order of magnitude brighter than what we measure. For the purposes of these estimates we assume a fiducial factor of $50$ such that $L_{X}=2.0\times10^{38}$\,erg\,s$^{-1}$ in Epoch~1 and $L_{X}=1.1\times10^{38}$\,erg\,s$^{-1}$ in Epoch~2. Bearing in mind that we only find an upper limit for $\log_{10} \xi$
in Epoch~1, we find a lower limit $r_{\rm Si}\gtrsim1.6\times10^{11}$\,cm for the Si-emitting medium, whilst in Epoch~2 we find $r_{\rm Si}\gtrsim1.3\times10^{11}$\,cm. For the Fe-emitting plasma, we estimate $r_{\rm Fe}\lesssim3.1\times10^9$\,cm and $r_{\rm Fe}\sim1.1\times10^{10}$\,cm in Epochs 1 and 2, repsectively - much closer to the BH than the Si-emitting medium. 

In addition, we can use the best-fit $n_{e}$ to estimate the size of the emitting regions, since $N_{e}=n_{e}\Delta r$, where $N_{e}$ is the column density of the photoionized plasma (which is also fit by the {\sc xstar} grid) and $\Delta r$ is the size of the medium. In Epoch~1 we find $N_{e}>2.6\times10^{21}$\,cm$^{-2}$ and $N_{e}=1.3^{+0.2}_{-0.9}\times10^{21}$\,cm$^{-2}$ for the Si- and Fe-emitting plasma, respectively, implying $\Delta r_{\rm Si}>3.9\times10^{8}$\,cm and $\Delta r_{\rm Fe}<1.4\times10^{8}$\,cm. In Epoch~2 we find ${N_{e}>2.8\times10^{21}}$\,cm$^{-2}$ and ${N_{e}<3.0\times10^{20}}$\,cm$^{-2}$ for the Si- and Fe-emitting plasma, respectively, implying ${\Delta r_{\rm Si}>5.5\times10^{8}}$\,cm and ${\Delta r_{\rm Fe}<4.3\times10^{7}}$\,cm. This implies that the Fe-emitting region is more compact than the Si-emitting plasma, in both \cha\ epochs.

\subsection{Line Origins: A spherical outflow?}

We find that the narrow lines in the two \cha/HETG spectra of \src\ are best described by two photoionized plasma components in conjunction with a partially covered disc blackbody model. Although the BB algorithm we used to perform the initial line search did not pick out any blue-shifted absorption features at 95\% confidence, we find that the best-fit twin photoionized plasma model implies the existence of weak \mbox{P-Cygni} profiles in the spectra (at least for the non-Fe plasma; see Fig. \ref{fig:Si_xstar}, red, and Fig. \ref{fig:full_xstar}), hinting at the existence of a spherical outflow, or at the very least, one with a non-zero solid angle \citep[see e.g.][]{Mauche-1987,Matthews-2015}. The best-fit velocities in Epoch~1 (from the $z$ values in Table \ref{tab:xstar_model}) are $v=-150^{+60}_{-90}$\,km\,s$^{-1}$ and $v=480\pm180$\,km\,s$^{-1}$ for the blue absorption and red emission components of the non-Fe plasma, respectively. In Epoch~2 we find $v=-480^{+240}_{-90}$\,km\,s$^{-1}$ and $v=240^{+120}_{-60}$\,km\,s$^{-1}$. Although these are lower velocity shifts than, say, those seen in V404\,Cyg \citep{King-2015}, similar centroid velocities were seen in H$\alpha$ during previous outbursts of \src\ \citep[see fig. 4 of][]{Munoz-Darias-2018}, who ascribe the observed profiles to a spherical wind. In addition, \citet{Lindstrom-2005} noted a weak P Cygni-like profile in Fe~{\sc ii} $\lambda5179$ during the 2004 outburst, with the absorption component blueshifted by $v\sim500$\,km\,s$^{-1}$, again consistent with our measurements in the second \cha\ epoch. We must emphasise here that, as the blue-shifted absorption features were not detected at 95\% confidence by the BB line search, the spherical wind hypothesis is speculative. However, such a hypothesis does line up with previous, optical, studies of \src.

\section{High-resolution X-ray spectroscopy with future instrumentation}
\label{sec:future}

Since launch, \cha/HETGS has proved a powerful tool to study astrophysical plasmas at high resolution. The next generation of X-ray instrumentation will probe X-ray spectra with resolutions as low as a few eV ($\sim10^{-4}$\,\AA). Here we present simulated spectra of \src\ to visualize how our knowledge of the spectrum might improve with the successful launch of future X-ray missions. In particular we focus on the Fe and Si regions.

The {\em X-ray Imaging and Spectroscopy Mission} \citep[{\em XRISM};][]{Tashiro-2018} will carry a microcalorimeter called Resolve \citep{Ishisaki-2018} as part of its instrumentation suite. Resolve's energy resolution is expected to be 5--7\,eV in the 0.3--12\,keV bandpass, with a nominal requirement of at least 7\,eV (0.0024\,\AA). The Fe~{\sc xxvi} Ly$\alpha$ line is actually a fine-structure doublet, whereas Fe~{\sc xxv} is a He-like triplet similar to the Si~{\sc xiii} lines studied in this work. These lines are unresolved in \cha/HETG,\footnote{Though see the insets in Fig. \ref{fig:Fe_region}} but should be resolved by {\em XRISM}/Resolve.

To test this, we performed simulations in the 1.7--1.9\,\AA\ range using the publicly available response matrices for the minimum energy resolution requirement of 7 eV. Our input spectrum included six Gaussians, two consist of the Ly$\alpha_{3/2}$ ($\lambda=1.778$\,\AA) and Ly$\alpha_{1/2}$ ($\lambda=1.783$\,\AA) doublet with a flux ratio of 2:1 as implied by the {\tt AtomDB}. The remaining 4 Gaussian lines make up the $r$ ($\lambda=1.850$\,\AA), $i$ ($\lambda_1=1.855$\,\AA, $\lambda_2=1.860$\,\AA) and $f$ ($\lambda=1.868$\,\AA) components of the He-like Fe~{\sc xxv}, which is actually a quadruplet due to the intercombination doublet. We use a fiducial flux ratio of $r:i_1:i_2:f=3:2:2:8$, as implied by the measured $R\sim2$ and $G\sim4$ values of Si~{\sc xiii} from the first \cha/HETG epoch, but note that this may not hold for He-like Fe~{\sc xxv} (see e.g. Fig. \ref{fig:Fe_region}; inset). The width of the lines was fixed to 0.0007\,\AA. We find that with just a 5\,ks exposure with {\em XRISM}/Resolve, the Ly$\alpha$ doublet can easily be resolved. In addition, we are able to resolve the $r$ and $f$ components of Fe~{\sc xxv}, though the two intercombination lines remain undetected.

\begin{figure}
    \centering
    \includegraphics[width=0.475\textwidth]{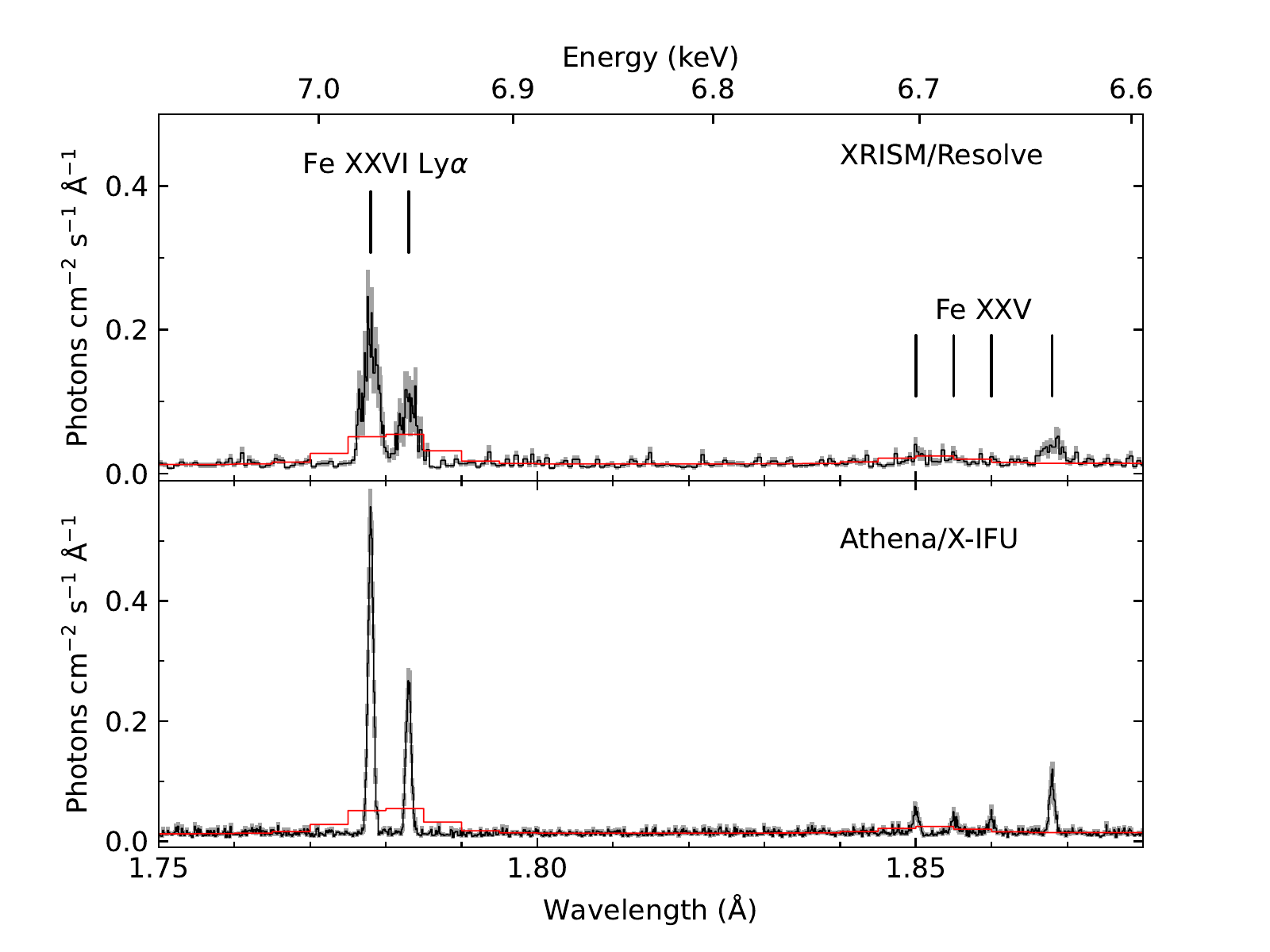}
    \caption{Simulations of the 1.7--1.9\,\AA\ region of the X-ray spectrum of \src. The top panel shows a simulated spectrum with {\em XRISM}/Resolve, while the bottom panel shows a simulation with {\em Athena}/X-IFU. Both spectra are simulated assuming a 5\,ks exposure time. In the top panel, the wavelengths of the Fe~{\sc xxvi} Ly$\alpha$ doublet and the Fe~{\sc xxv} He-like lines - which are actually resolved as a quadruplet by {\em Athena} - are labeled. In both panels, the red line represents the best fit model to the Epoch~1 \cha/HETG spectrum of the Fe region (see Fig. \ref{fig:Fe_region}). Both simulated spectra are grouped such that each bin has a minimum of 10 counts.}
    \label{fig:spec_sims}
\end{figure}

Beyond {\em XRISM}, the {\em Advanced Telescope for High Energy Astrophysics} \citep[{\em Athena};][]{Nandra-2013,Barret-2020} has a target launch date of 2035 and will carry the X-ray Integral Field Unit \citep[X-IFU][]{Barret-2016}. X-IFU will\,provide better than 2.5\,eV ($\sim0.0006$\,\AA) resolution up to 7\,keV, with a large effective area ($\sim970$\,cm$^2$ at 7\,keV). We simulated {\em Athena}/X-IFU spectra of the Fe region using the same input model as for {\em XRISM}/Resolve, though with the line widths instead fixed to $0.0002$\,\AA. We find that we are able to easily resolve the Fe~{\sc xxvi} Ly$\alpha$ doublet, as well as Fe~{\sc xxv} $r$ and $f$ in just 1\,ks. Increasing the exposure time to 5\,ks, as we did with the {\em XRISM}/Resolve simulation, allows us to measure the individual components of the intercombination doublet. The two simulated spectra from {\em XRISM}/Resolve and {\em Athena}/X-IFU are presented in Fig. \ref{fig:spec_sims}. 

Resolving the He-like lines of Fe will allow us to perform plasma diagnostics on the iron-emitting plasma much like we did in Section \ref{sec:diagnostics} with Si~{\sc xiii}. Studying the Fe~{\sc xxv} complex will allow us to probe and accurately quantify higher densities than those allowed by lower $Z$ ions such as Si~{\sc xiii}, up to a few $10^{17}$\,cm$^{-3}$, as well as probe higher temperatures via the ${\rm G}$ ratio \citep{Porquet-2010}. Furthermore, the large collecting area of the future X-ray missions, combined with the increase in spectral resolution, will enable us to trace line properties on $\sim$ks timescales, measuring changes in, for example, equivalent widths over the binary orbital period, opening up the ability to study the spatial distribution of the line-emitting material at different orbital phases (Doppler tomography).

The large effective area of {\em Athena}/X-IFU will also allow us to probe plasma diagnostics at longer wavelengths, for example the Si~{\sc xiii} complex. We again simulated spectra of \src, this time in the Si region, using an input ratio $r:i_1:i_2:f=3:2:2:8$ for Si~{\sc xiii} as implied by the measured ${\rm R}\sim2$ and ${\rm G}\sim4$ from Epoch~1 of \cha/HETG. We acknowledge the large scatter in the measured ${\rm R}$ and ${\rm G}$ ratios (Si~{\sc xiii} $r$ is actually consistent with zero in the \cha\ data), but choose these ratios as an example to test the capabilities of {\em Athena}/X-IFU. In a 5\,ks exposure, we find that, though the Si~{\sc xiv} Ly$\alpha$ and Si~{\sc xiii} intercombination doublets remain unresolved, the triplet is easily detected. Utilizing the {\tt triplet\_fit} function, as we did in Section \ref{sec:diagnostics}, we measure ${\rm R}=2.2^{+0.3}_{-0.4}$ and ${\rm G}=3.5^{+0.7}_{-0.4}$. 

\section{Conclusions}
\label{sec:conclusions}

We have presented a study of the 2020 outburst of the BH-LMXB \src. We obtained spectra of the system with \nus\ in January 2020 when it was in Sun constraint for the majority of other telescopes. \nus\ revealed a hot, disc-dominated spectrum with a strong emission feature in the 6--7\,keV range. Two epochs of spectroscopy with the \cha/HETGS also found a similar X-ray continuum. The measured inner disc temperatures and luminosities in both epochs imply that the central engine of \src\ is obscured, and that X-rays are scattered above the disc.

\cha/HETG resolved the 6--7\,keV emission feature seen by \nus\ into two strong, highly ionized Fe lines at 6.7 and 6.97\,keV (1.855 and 1.780\,\AA, respectively). \cha\ also revealed a number of other highly-ionized metal emission lines, including Si~{\sc xiv} Ly$\alpha$ and even He-like Si~{\sc xiii}. We used the He-like Si lines to perform some plasma diagnostics, measuring the ratio between the individual triplet components to find that the emission was consistent with a photoionized plasma. We then generated grids of photoionized plasmas using {\sc xstar}, finding that the \cha\ spectra in both epochs could be well-described by two photoionized components, one to reproduce the Fe lines and a second for the non-Fe metals. This implies that there were two distinct emission regions responsible for the narrow line emission seen in \src, with separate ionization parameters and densities for each.

Though \cha/HETGS revealed a myriad of strong lines in the X-ray spectrum of \src, we show that the next-generation of X-ray instrumentation will deliver higher S/N spectra in a fraction of the time. In particular, {\em Athena}/X-IFU will provide high spectral resolution and effective area such that we will be able to track X-ray line properties over meaningful timescales, such as the binary orbital period, in BH-LMXBs such as \src.

\section*{Acknowledgements}
We thank the anonymous referee for their insightful comments that helped improve this manuscript. This research has made use of ISIS functions (ISISscripts) provided by ECAP/Remeis observatory and MIT (\href{http://www.sternwarte.uni-erlangen.de/isis/}{http://www.sternwarte.uni-erlangen.de/isis/}). AWS would like to thank the \nus\ Galactic Binaries working group for useful discussions regarding this work. Support for this work was provided by the National Aeronautics and Space Administration through Chandra Award Number GO0-21128X issued by the Chandra X-ray Center, which is operated by the Smithsonian Astrophysical Observatory for and on behalf of the National Aeronautics Space Administration under contract NAS8-03060. COH is  supported by NSERC Discovery Grant RGPIN-2016-04602; GRS is supported by RGPIN-2016-06569 and RGPIN-2021-0400.

\section*{Data Availability}

The data underlying this article are publicly available and can be accessed via the HEASARC at \href{https://heasarc.gsfc.nasa.gov/docs/archive.html}{https://heasarc.gsfc.nasa.gov/docs/archive.html}.



\bibliographystyle{mnras}
\bibliography{V4641Sgr.arxiv.bib} 








\bsp	
\label{lastpage}
\end{document}